\documentclass[trackchanges, twocolumn]{aastex701}
\usepackage{graphicx}
\usepackage{subfigure}
\usepackage[utf8]{inputenc}
\usepackage{booktabs} 
\usepackage{array}    
\usepackage{hyperref}

\usepackage{amsmath}
\usepackage{xcolor}
\usepackage{xspace}

\newcommand{\apsr}{PSR~J1840+1102\xspace}

\newcommand{\ppcc}{$\,$pc$\,$cm$^{-3}$\xspace} 

\begin{document}

\title{Survey of Compact Sources for Pulsars and Exotic Objects -- II.\\ Precision Timing of Millisecond Pulsar J1840+1102}
 
\author[orcid=0009-0001-8761-0498,gname='Ayushi', sname='Shrivastava']{Ayushi Shrivastava}
\affiliation{National Centre for Radio Astrophysics (NCRA - TIFR), Pune - 411007, India}
\email[show]{ashrivastava@ncra.tifr.res.in}  

\author[orcid=0000-0002-0862-6062,gname=Yogesh, sname='Maan']{Yogesh Maan} 
\affiliation{National Centre for Radio Astrophysics (NCRA - TIFR), Pune - 411007, India}
\email{ymaan@ncra.tifr.res.in}

\author[0000-0003-4528-2745,gname=M. A.,sname=Krishnakumar]{M. A. Krishnakumar}
\affiliation{National Centre for Radio Astrophysics (NCRA - TIFR), Pune - 411007, India}
\affiliation{Radio Astronomy Centre, NCRA-TIFR, Udagamandalam, India}
\email{ kkma@ncra.tifr.res.in}

\author[orcid=0009-0000-6605-3162,gname='Banshi',sname='Lal']{Banshi Lal}
\affiliation{National Centre for Radio Astrophysics (NCRA - TIFR), Pune - 411007, India}
\email{blal@ncra.tifr.res.in}

\author[0000-0001-8503-6958,gname=Joeri,sname=van Leeuwen]{Joeri van Leeuwen}
\affiliation{ASTRON, The Netherlands Institute for Radio Astronomy, Oude Hoogeveensedijk 4, 7991 PD Dwingeloo, The Netherlands}
\affiliation{Leiden Observatory, Leiden University, Einsteinweg 55, 2333 CC Leiden, The Netherlands}
\affiliation{Leiden Institute for Advanced Computer Science (LIACS), Leiden University, Einsteinweg 55, 2333 CC Leiden, The Netherlands}
\email{leeuwen@astron.nl}

\begin{abstract}
We present the timing analysis of the eclipsing millisecond pulsar \apsr, discovered in the Survey of COmpact sources for Pulsars and Exotic objects (SCOPE) using the Giant Metrewave Radio Telescope (GMRT). \apsr has a short spin period of  1.60\,ms and an orbital period of 1.68\,days, with a dispersion measure (DM) of 98.8 \ppcc. We report timing analysis over a 2.7-year baseline, yielding a phase-connected solution with a post-fit residual RMS of 2.1\,$\mu$s. Enabled by the timing solution, we report discovery of pulsed gamma-rays from \apsr, with the gamma-ray peak aligned with that of the radio profile. \apsr also exhibits remarkably short eclipses, lasting only around $0.4\%$ of the total orbit at 650\,MHz. Based on optical searches, the orbital parameters, short eclipses, and comparison with the known binary pulsar population, we infer a Helium white dwarf (He-WD) surrounded by plasma to be the most likely companion. Furthermore, using GMRT observations covering 300 to 750\,MHz, we present a study of the interstellar medium properties towards \apsr including refractive scintillation, DM variations and scattering, and report a very low scattering index of $1.65\pm0.15$. The low timing residuals and lack of any prominent orbital parameters or DM variations support the overall suitability of \apsr for inclusion in the pulsar timing array campaigns.
\end{abstract}


\section{Introduction}\label{sec:intro}

Pulsar timing involves measuring a sequence of pulse times of arrival (ToAs) from a pulsar over extended durations and fitting a comprehensive model that describes the pulsar's spin, astrometry, and, if applicable, binary parameters. When well-modeled, pulsars serve as powerful laboratories to probe extreme physical environments, investigate the interstellar medium, and test general relativity in the strong-field regime \citep{lorimer_handbook_2004}. Particularly for millisecond pulsars (MSPs) in binary systems, timing enables precise measurements of the neutron star (NS) and companion masses, proper motions, and orbital dynamics, which help decode the stellar evolution and binary interactions. High-precision timing studies also allow probes of the equation of state of cold nuclear matter through accurate NS mass estimates \citep[e.g.,][]{Clark_2023}. 

\par 
Multi-wavelength observations often complement the timing studies. Radio emission from pulsars originates from charged bundles moving relativistically along curved open dipolar magnetic field lines \citep{1975RFM}. The observed optical emission often arises from the companion star, which helps in characterizing the type of companion, inclination angle of the orbit, and, in some cases, constraining the mass of both bodies in the system. Many pulsars also exhibit X-ray emission through polar cap heating and gamma-ray emission via curvature radiation. Detailed timing studies, combined with such multi-wavelength data, offer a deeper understanding of emission mechanisms, companion type, and evolutionary history of the system.

\par 
Furthermore, a major collaborative effort within the pulsar community is the formation of Pulsar Timing Arrays (PTAs) --- networks of highly stable MSPs monitored for correlated timing deviations \citep{1983ApJ...265L..39H}. These efforts aim to detect ultra-low-frequency gravitational waves, primarily originating from supermassive black hole binaries. Recent results from global PTA collaborations suggest early evidence of a stochastic gravitational wave background, marking a major milestone in observational astrophysics \citep{2023aAgazie_NANOGrav, 2023Reardon_PPTA, 2024Antoniadis_EPTA_InPTA}.

\par 
Given these far-reaching applications, it is essential to continue searching and timing new millisecond pulsars. Numerous surveys, both blind and targeted, are actively observing to expand the known pulsar population. Out of $\sim$ 4400 pulsars currently known, 14$\%$ reside in binary systems.


\par
In the ongoing Survey of COmpact sources for Pulsars and Exotic objects \citep[SCOPE\footnote{\url{http://www.ncra.tifr.res.in/~ymaan/scope.html}};][]{2026SCOPE}, using the upgraded Giant Metrewave Radio Telescope (uGMRT) \citep{Swarup-GMRT, YG-GMRT} at band-4 (550$-$750\,MHz), we recently discovered an eclipsing MSP, PSR~J1840+1102. The discovery observation suggested that this pulsar resides in a binary system and exhibits an extremely short spin period (P) of 1.60\,ms, at a dispersion measure (DM) of 98.8\,\ppcc. In this work, we present a detailed study of this recently discovered pulsar, including a timing solution using observations over more than 2 years, multi-wavelength probes of the system, and the interstellar medium (ISM) properties towards the pulsar.

\par 
The remaining part of the paper is organized as follows. In Section~\ref{sec:obs}, we provide details of the radio observations of PSR J1840+1102. Section~\ref{sec:data} describes the data reduction and different analysis techniques employed to derive the timing ephemeris, correct for the varying DM, measure the eclipse duration, search for multi-wavelength counterparts, and characterize the ISM effects. In Section~\ref{sec:discussion}, we provide a detailed discussion of our results, followed by conclusions in Section~\ref{sec:conclusion}.

\section{Observations}\label{sec:obs}

The flux density estimates for \apsr from SCOPE images combined with those from TGSS \citep{2014TGSS,TGSS1} and RACS \citep{McConnell_2020_racs}, suggest the pulsar to have a steep spectrum with spectral index $-2.8\pm0.5$ \citep{2026SCOPE}. Balancing the steep spectrum against the frequency-dependent effects such as the interstellar scattering of the pulse profile, band-4 of uGMRT provides the optimal observing frequency range for regular timing of this source. In addition to conducting regular timing observations in band-4, we also conducted a few observations in band-3 (300$-$500\,MHz) of GMRT to investigate the frequency evolution of the pulse profile.

\par 
This work uses band-4 timing observations conducted between June 2023 and March 2026. Each observation had a typical on-source integration time of about 30 minutes, using a 200\,MHz bandwidth centered at 650\,MHz. Data were recorded using the coherently dedispersed phased-array (CDPA) beam, wherein the spectral voltages in each frequency sub-band are corrected in real-time for dispersive delays at the nominal DM value 98.8\,\ppcc. As high temporal resolution is critical for obtaining a good timing solution, the band-4 observations were recorded with a sampling time of 5.12\,$\mu$s. At band-3 (300$-$500\,MHz), partial smearing of the pulse profile due to scatter-broadening is unavoidable, and thus, we adopted a sampling time of 10.24\,$\mu$s for the band-3 observations.

\par We strategically distributed our initial timing observations at a higher cadence, with separations ranging from a day to about a week, which helped constrain the orbital parameters and establish the initial phase connection. The subsequent observations were conducted with a more uniform cadence, with gaps ranging between 2 and 3 weeks. 

\section{Data Analysis and Results}\label{sec:data}

Radio observations are often heavily contaminated by both narrow-band and broad-band, short-lived, man-made signals, known as radio frequency interference (RFI). We performed RFI excision on the data from the individual sessions using the mitigation tool \texttt{RFIClean}\footnote{\url{https://github.com/ymaan4/RFIClean}} \citep{maan_fourier_2021}. In addition to RFI excision, \texttt{RFIClean} also converts the native GMRT data to SIGPROC filterbank format. The RFI-excised filterbank files were then subjected to \texttt{digifil} from \texttt{DSPSR}, to down-sample from 16\,bits to 8\,bits per sample, as well as normalize the effects of the varying gain of the band-pass filter across the observed bandwidth. A second round of RFI excision was then performed using \texttt{rfifind} from the pulsar exploration and search toolkit \citep[\textsc{PRESTO};][]{PRESTO}.

\par As the Earth's local observing frame is non-inertial due to its rotation and orbital motion, all measurements must be transformed to the centre of mass of the solar system, known as the Solar System Barycenter (SSB) frame. The quantities measured in the barycentric frame can seamlessly be combined and compared with data across different observing epochs and from different observatories. We used \texttt{prepdata} from \textsc{PRESTO}, to incoherently dedisperse the filterbank data from individual sessions using a DM value of 98.8\,\ppcc as well as barycenter the data, and prepare a band-averaged, dedispersed, barycentered time series. A Fourier transform was applied to the time series using \texttt{realfft}. For the \apsr system, the line-of-sight (LOS) acceleration, measured from the discovery observation as well as from the subsequent ones, is well within the range of typical binary MSPs. To account for the effect of the binary motion while searching for the pulsar's periodic signal, we conducted an acceleration search on the \texttt{realfft} output using \texttt{accelsearch}. We also employed harmonic summing using \texttt{accelsearch}. Once the \apsr was identified in the candidate list, we folded the candidate using \texttt{prepfold} and estimated the best barycentric period, $P$, and period derivative, $\dot{P}$. We note that \texttt{realfft}, \texttt{accelsearch} as well as \texttt{prepfold} are a part of \textsc{PRESTO}, and the above procedure was followed for the data from individual sessions. 

\subsection{Initial Orbital Solution}

\begin{figure}
    \centering
    \includegraphics[width=0.47\textwidth]{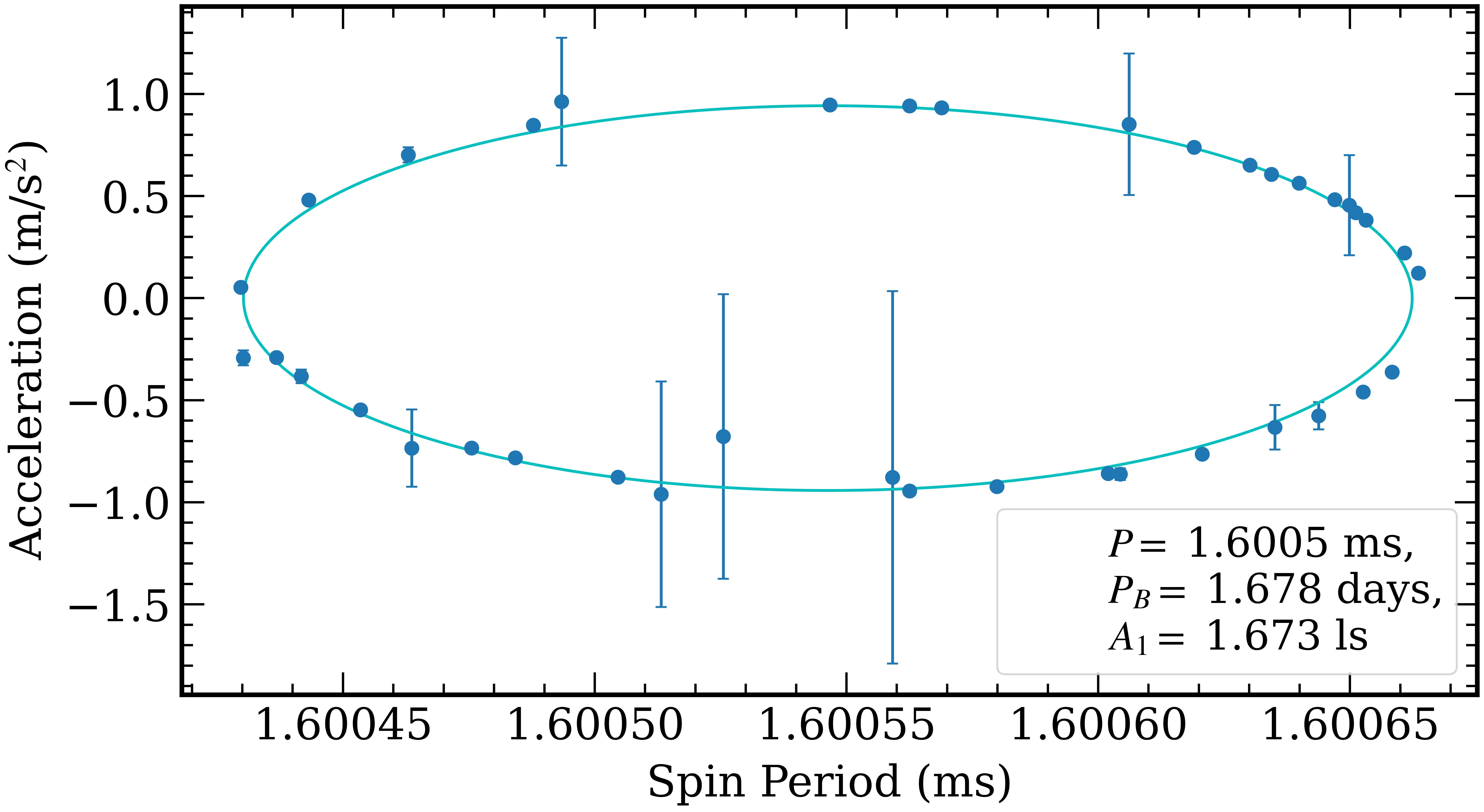}
    \caption{Orbital parameter estimation for PSR J1840+1102 using the ellipse-fitting technique. The plot displays the measured line-of-sight acceleration ($a_{LOS} = c\dot{P}/P$) as a function of the observed spin period across multiple observing epochs. The measured data points (blue) clearly trace a closed ellipse (solid curve, fitted), confirming the presence of a nearly circular binary orbit ($e \simeq 0$). The parameters extracted using this analysis are listed in the figure legends.}
    \label{fig:ellipse}
\end{figure}

To obtain the initial estimate of the binary orbital period, $P_B$, we used the ellipse fitting method described by \cite{freire_determination_2001}. This technique uses the fact that plotting $\dot{P}$ versus $P$ maps out a closed loop. If we further simplify the equations assuming a circular orbit, i.e., eccentricity $e$, as well as the longitude of periastron $\omega$, are 0, the parametric equations for the observed spin period and acceleration should trace a perfect ellipse in the acceleration ($a_{LOS} = c\dot{P}/P$) vs $P$ plane. 

\par
By estimating $a_{LOS}$ from our measured $P$ and $\dot{P}$ values, we see that the data points indeed trace an ellipse in the $a_{LOS}$ vs. $P$ plane, providing an initial confirmation that the orbit is nearly circular (see Figure \ref{fig:ellipse}). It should be noted that a few data points exhibit significantly larger uncertainties in acceleration, primarily due to shorter integration times at those specific epochs. We initially used the publicly available script \texttt{estimateOrbit.py} from \texttt{CandyCracker}\footnote{\url{github.com/mcbernadich/CandyCracker}} to fit an ellipse to our above measurements. From this fit, we extracted initial estimates for both $P_B$ and the projected semi-major axis, $A_1$. 

\par Building on the above initial estimates, we employed \texttt{fit-circular-orbit.py} and \texttt{fitorb.py} from \textsc{PRESTO}, to also obtain the initial estimates for the epoch of passage through periastron, $T_0$, and the longitude of periastron, $\omega$, in addition to better estimates for $P_B$ and $A_1$. These routines assume a Keplerian orbital model and evaluate how the observed spin period ($P_{obs}$) varies at different orbital phases due to the Doppler effect. 

\subsection{Timing}

\subsubsection{Phase-connection}

Phase-connection is the process of accounting for every single pulsar rotation between sequential times of arrival (ToAs) of the pulses estimated from observations. Phase connection is the key requirement for obtaining a reliable timing model. The first step in achieving a phase-connected timing solution is to estimate pulse ToAs. To do so, we first partially folded the filterbank files from the individual sessions using \textsc{dspsr} \citep{2004PSRCHIVE_PSRFITS, 2011van_dspsr} software and the corresponding best period and period derivative estimated using \texttt{accelsearch}. The output from this step is a partially folded \texttt{PSRFITS} format file, containing 20\, second-long sub-integrations for typical observing duration of 30 minutes for each of the sessions. 
\par
Next, to generate a template of the pulsar's average profile, we used the frequency- and time-averaged folded profile from an observation that yielded the highest Signal-to-Noise Ratio (SNR). The noise-less representation of the average profile is obtained by fitting a number of Gaussian components to the average profile using \texttt{paas} from \textsc{PSRCHIVE} \citep{2012PSRCHIVE}. The best fit to the average profile is used as a template. The ToAs are estimated by cross-correlating a given pulse profile with the noise-less template. To obtain the initial phase connection, it is important to generate multiple ToAs across the individual observations. After fully averaging over the frequency bandwidth, we further averaged every 16 consecutive sub-integrations (amounting to an averaging over nearly 5 minutes). These fully band-averaged and about 5\,minutes sub-integrated profiles were then used to obtain the ToAs, using \texttt{pat} from \texttt{PSRCHIVE}. Starting with the preliminary ephemeris obtained earlier and using the dense observing cadence of the initial epochs, a phase-connected solution was obtained using \texttt{tempo2} \citep{edwards_tempo2_2006}, using the \texttt{BT76} binary model \citep{blandford_arrival-time_1976}.

\par 
After obtaining a phase-connected solution, we also tried \texttt{DD86} binary model \citep{damour_general_1986}, which includes treatment of short-term periodic quantities, and fits for Shapiro and aberration delays. \texttt{BT76} and \texttt{DD86}, both resulted in nearly the same post-fit residuals, with an rms of $\simeq3\,\mu s$, for a timing baseline of 1.2 years.

\par
The minimum rms of the post-fit residuals was achieved using the \texttt{ELL1} binary model \citep{lange_precision_2001}. This model works best for nearly circular orbits, and avoids covariance between $T_0$ and $\omega$. Instead of fitting $T_0$, it fits for the time of passage through the ascending node, $T_{asc}$, which remains well-defined regardless of the orbital eccentricity. $T_{asc}$ is related to $T_0$ and $\omega$ in the following way.

\begin{equation}\label{eqn:2}
    T_{asc}\: =\:\left(T_0 - \frac{\omega P_B}{2\pi}\right)
\end{equation}

Furthermore, in nearly circular orbits, $T_0$ is not a prominent feature in ToAs. \texttt{BT76} and \texttt{DD86} models do not take this fact into account, and it leads to very high covariance between $\omega$ and T$_0$ while fitting. Moreover, rather than fitting for the eccentricity directly, \texttt{ELL1} uses two Laplace-Lagrange parameters, $\epsilon_1=e\sin(\omega)$ and $\epsilon_2=e\cos(\omega)$. Fitting for these three parameters effectively breaks the degeneracy and provides a stable and reliable fit for circular systems.

\par Next, we used wavelet transform for template generation using the highest SNR profile. Wavelet transform smooths out the noisy features and removes any subjective choices that the previous template generation method could have been affected. This is achieved by using the \texttt{WaveletSmooth} implementation in PSRCHIVE. We adopted the wavelet whose smoothed profile gave the lowest RMS residual with respect to the original data. Using this new template, and observations up to March 2026, we extended the timing baseline to nearly 2.7\,years and achieved a post-fit residual RMS of 2.5\,$\mu$s. We note that the above longer baseline also allowed us to fit for and estimate the proper motion parameters.

\begin{figure*}
    \centering
    \includegraphics[width=0.9\textwidth]{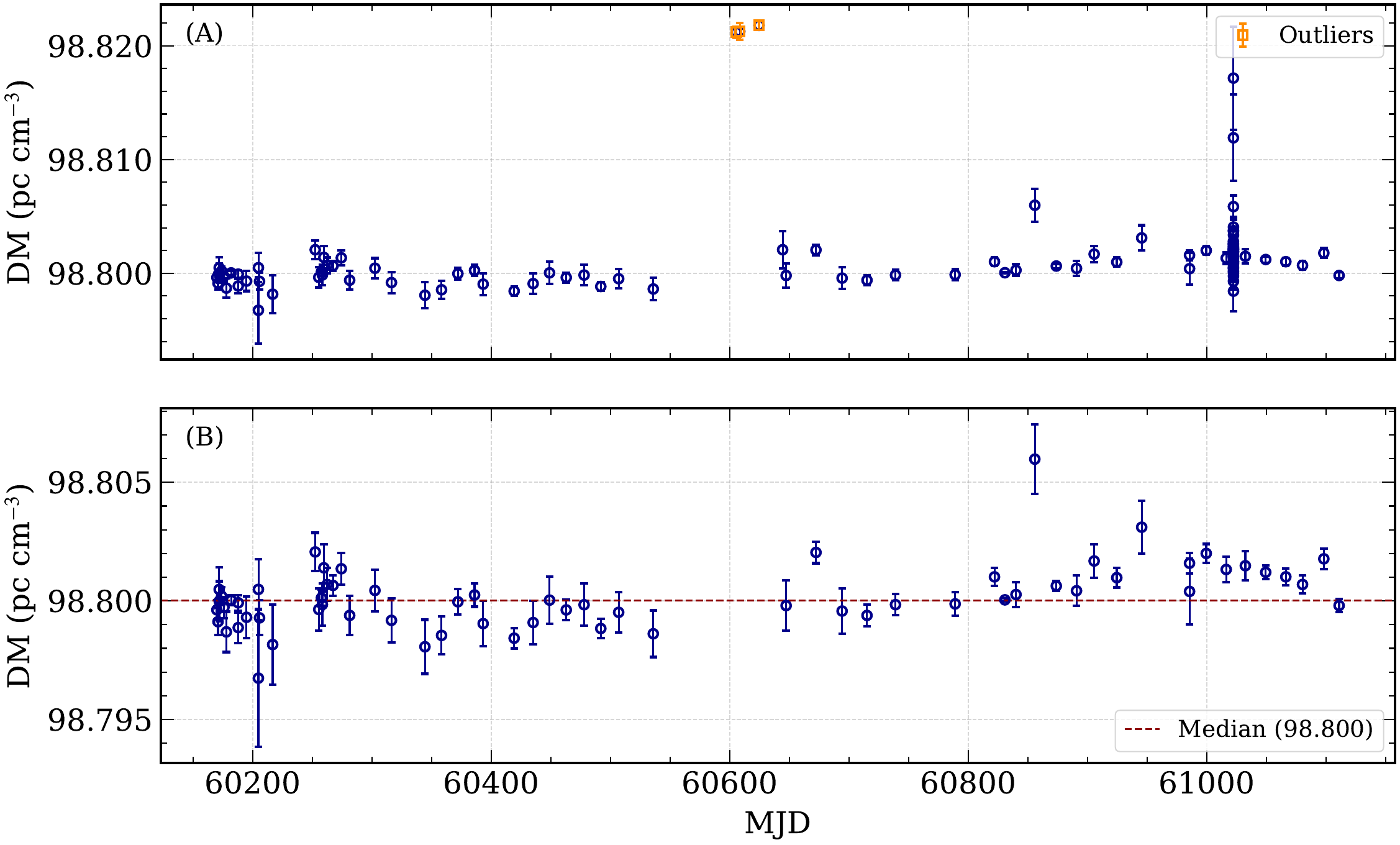}
    \caption{Temporal evolution of the DM for \apsr, spanning a $\sim$2.7 year observing baseline. The top panel (A) shows the measured DM for all the observations. The bottom panel (B) shows the same measurements but excluding the ones around the eclipse and the three outliers marked as orange squares. The red dashed line marks the median DM, again excluding the outliers and those around the eclipse.}
    \label{fig:DMT_mjd}
\end{figure*}

\subsubsection{Frequency resolved timing and DM Analysis}\label{sec:dm_analysis}

To further improve the timing solution, we measured and corrected for the temporal, epoch-to-epoch variations in DM. We estimated the optimal DM for each observing epoch using \texttt{dmcalc.py} \citep{MAK_2021}, which uses \texttt{tempo2} as its backend fitting engine. For this purpose, we generated a frequency-resolved template via wavelet smoothing from a high SNR observed profile, by dividing the observing bandwidth into 16 sub-bands and fully integrating over time. Using this frequency-resolved template, frequency-resolved ToAs were extracted from each session, and fitting for DM was performed for each epoch.

\par
\begin{figure*}
    \centering
    \includegraphics[width=0.8\textwidth]{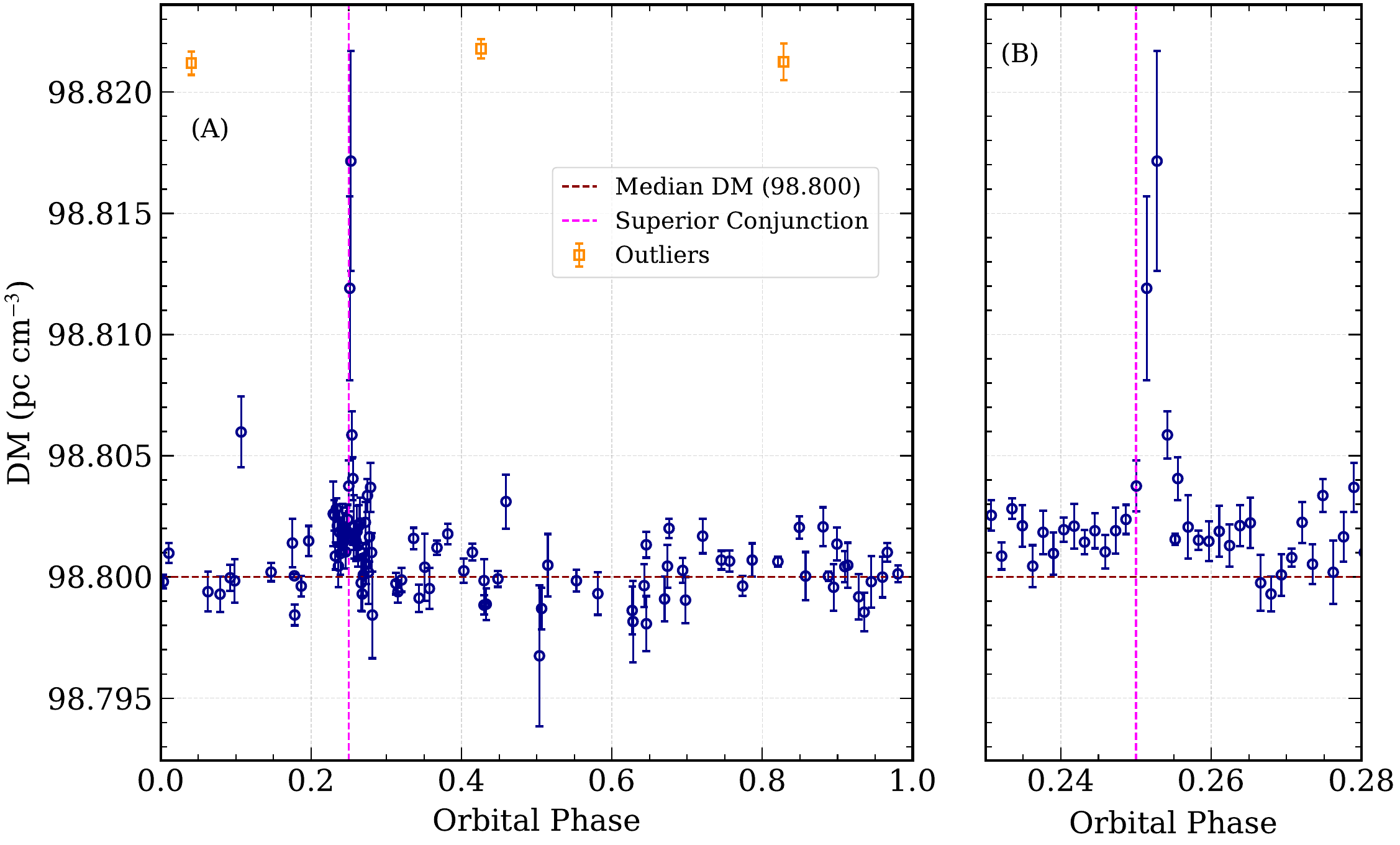}
    \caption{DM variations, same as those presented in Figure~\ref{fig:DMT_mjd}, but now shown as a function of orbital phase. The left panel (A) shows the full orbital phase range, while the right panel (B) presents a zoom-in around the superior conjunction. The horizontal red dashed lines in both panels represent the median DM, as in Figure~\ref{fig:DMT_mjd}.}
    \label{fig:DMT_Phase}
\end{figure*}

 For each of the epochs, we estimated a single, average DM over the entire, typically 30 minutes long, observation. However, one of our observation is 2\,hours long, which was aimed to cover the orbital range 0.2$-$0.3 to characterize eclipses. To map the highly dynamic DM variations during and around the eclipse, we divided this observation into shorter, nearly 3.5-minute-long segments (i.e., each segment covering approximately 0.15$\%$ of the total orbit), using \texttt{psrsplit} from \texttt{PSRCHIVE}. This high-resolution splitting helped in resolving the expected DM excess peaking at the superior conjunction, i.e., at the orbital phase of 0.25. Figures~\ref{fig:DMT_mjd} and \ref{fig:DMT_Phase} show the above deduced DM variations as a function of the observing epoch and the orbital phase, respectively.

\par 
We detect isolated, anomalously high DM values during a few specific epochs, significantly away from the orbital phase of 0.25. To determine if these high DM values were caused by solar wind density fluctuations, we cross-referenced the observation dates with records of solar activity, but no corresponding major flares were found. The exact origin of these anomalous measurements is not clear. Apart from the above anomalous values and those caused by the intra-binary material around the eclipse, we observe only slow, long-term variations in DM across the timing baseline, without any systematic trends across the orbital phase (see Figures~\ref{fig:DMT_mjd} and \ref{fig:DMT_Phase}).

\begin{figure*}
    \centering
    \includegraphics[width=0.8\textwidth]{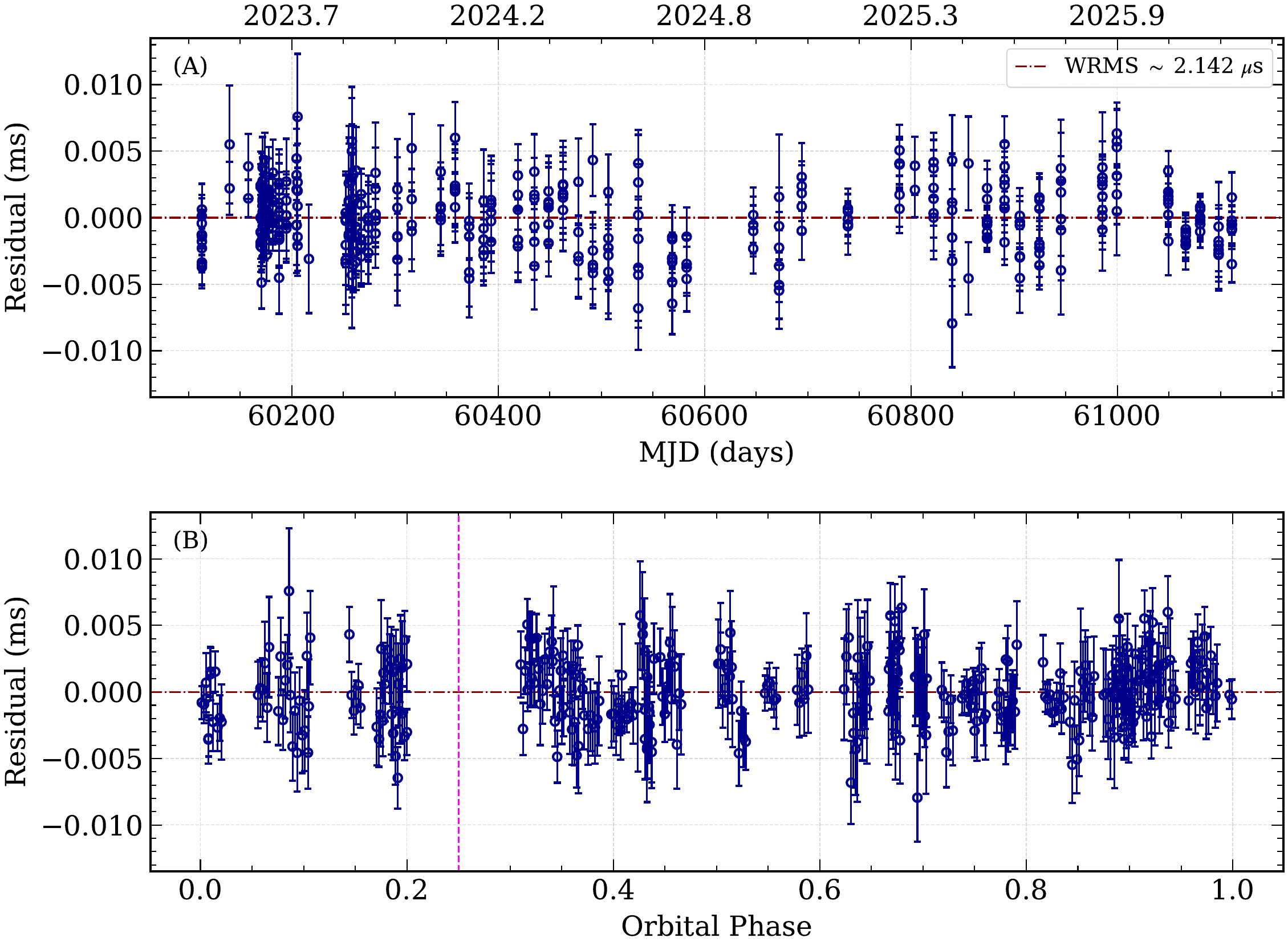}
    \caption{Frequency-averaged timing residuals derived from the ELL1 model after correcting for DM separately for each epoch. The top panel (A) shows the post-fit residuals as a function of epoch, while the bottom panel (B) shows them as a function of the orbital phase. All the ToAs shown here are from band-4 observations. ToAs between 0.2 and 0.3 orbital phase, those corresponding to the three DM outliers mentioned in the text (also see Figures~\ref{fig:DMT_mjd} and \ref{fig:DMT_Phase}), and ToAs with error $>5\mu$s, are excluded before fitting. The achieved weighted RMS is 2.142\,$\mu$s.}
    \label{fig:freq_average}
\end{figure*}

\subsubsection{DM-corrected timing}

\par
We applied the above epoch-specific DM corrections to the \texttt{PSRFITS} data archives using \texttt{psredit} from \textsc{PSRCHIVE}, and regenerated frequency-averaged ToAs. Correction for DM variations provided improvement to the timing solution, now with a post-fit residual RMS of $\sim$2.15\,$\mu$s (see Figure \ref{fig:freq_average}). This significant reduction in RMS indicates the presence of unmodeled dispersive delays in our earlier timing solution.

\par 
We note that the pulse profiles, and hence the ToAs, might be significantly perturbed by the eclipsing medium. To prevent our timing model from being skewed by such systematic delays associated with the eclipse, we excluded all the ToAs between the orbital phase range 0.2 to 0.3 from the above fitting procedure. Furthermore, to maintain the quality of the fit, we flagged any ToAs with uncertainties exceeding 5\,$\mu$s. Typically, \texttt{tempo2} fits never result in reduced $\chi^2$ = 1. This can happen due to the presence of unmodeled instrumental or astrophysical noise and incorrect uncertainties on the ToAs. To account for such factors, we used EFAC in the ToAs list to scale the uncertainties to the values that make reduced $\chi^2$ = 1, and are perhaps more consistent with the actual values.

We list all the fitted parameters as well as derived parameters for this system in Table~\ref{table:parameters}. 

\begin{table*}
    \centering
    \caption{Timing and orbital parameters for \apsr (parameter file is available on the SCOPE webpage\footnote{\url{https://www.ncra.tifr.res.in/~ymaan/scope.html}}).}
    \small 
    \begin{tabular*}{\linewidth}{l @{\extracolsep{\fill}} r}
        \toprule
        \multicolumn{2}{l}{\textbf{Observation and data-set parameters}} \\
        \midrule
        Pulsar name \dotfill & PSR J1840$+$1102 \\
        MJD range \dotfill & 60112.82 --- 61110.97 \\
        Total time span (year) \dotfill & 2.733 \\
        Number of ToAs \dotfill & 368 \\
        Epoch of frequency determination (MJD) \dotfill & 60506.79 \\
        Epoch of position determination (MJD) \dotfill & 60506.79 \\
        Epoch of dispersion measure determination (MJD) \dotfill & 60506.79\\
        Timing Model \dotfill & ELL1\\
        EFAC \dotfill & 1.2889\\
        Reduced Chi-square, $\chi^2$ \dotfill & 1.08769 \\
        Post-fit residual rms ($\mu$s) \dotfill & 2.142 \\
        \midrule
        \multicolumn{2}{l}{\textbf{Measured Quantities}} \\
        \midrule
        Right ascension (J2000), $\alpha$ \dotfill & 18:40:09.4336593(61) \\
        Declination (J2000), $\delta$ \dotfill & $+$11:02:09.55669(15) \\
        Dispersion Measure, DM (pc cm$^{-3}$) \dotfill & 98.8 \\
        Spin frequency, $\nu$ (Hz) \dotfill & 624.7862114899211(51) \\
        Spin frequency derivative, $\dot{\nu}$ (Hz s$^{-1}$) \dotfill & $-4.55844(41) \times 10^{-15}$ \\
        Orbital period, $P_{\mathrm{orb}}$ (days) \dotfill & 1.67988890549(14) \\
        Projected semimajor axis, $A_1$ (lt-s) \dotfill & 1.69902219(18) \\
        Time of ascending node, $T_{\mathrm{asc}}$ (MJD) \dotfill & 60072.787868705(48) \\
        EPS1, $\epsilon \sin({\omega}^*)$ \dotfill & $1.81(21) \times 10^{-6}$ \\
        EPS2, $\epsilon \cos(\omega)$ \dotfill & $ -0.33(21) \times 10^{-6}$ \\
        PMRA, mas yr$^{-1}$ \dotfill & $-$1.73(10) \\
        PMDEC, mas yr$^{-1}$ \dotfill & $-$6.31(22) \\
        Mass Function ($M_{\odot}$)\dotfill & 0.00186603057(46)\\
        Minimum, median and maximum companion mass, at i = 90$^\circ$, 60$^\circ$, 26$^\circ$  ($M_{\odot}$) \dotfill & 0.1622, 0.1895, 0.4120\\
        \midrule
        \multicolumn{2}{l}{\textbf{Conversion from ELL1 parameters}} \\
        \midrule
        Eccentricity \dotfill & $1.84(21) \times 10^{-6}$\\
        $^*$Longitude of periastron, $\omega$ (degrees) \dotfill & 100.3(6)\\
        Time of passage through periastron, $T_0$ (MJD) \dotfill & 60073.255(30)\\
        \midrule
        \multicolumn{2}{l}{\textbf{Derived Quantities}} \\
        \midrule
        DM distance YMW16 $|$ NE2025 model, $d$ (kpc) \dotfill & 5.60 $|$ 5.50 \\
        Total Proper motion; $\mu$ (mas/yr) \dotfill & 6.54(20) \\
        Shklovskii contribution $\dot{P}_{\rm shk}$  $\times 10^{-20}$ (s\,s$^{-1}$)$^{\dagger}$ \dotfill & $0.091$ \\
        Spin-down Luminosity, $\dot{E}$ $|$ $\dot{E}_{\rm shk}$ $\times 10^{35}$ (erg s$^{-1}$) \dotfill & 1.126 $|$ 1.04\\
        Characteristic age, $\tau_c$ $|$ $\tau_{\rm c,shk}$ (Gyr) \dotfill & 2.17 $|$ 2.36 \\
        Surface magnetic field, $B_s$ $|$ $B_{s,shk}$  $\times 10^8$ (G) \dotfill & 1.38 $|$ 1.33\\
        Magnetic Field at Light-Cylinder, B$_{LC}$ $|$ B$_{LC,shk}$ $\times 10^5$ (G) \dotfill & 3.06 $|$ 2.94\\
        \bottomrule
    \end{tabular*}
    
    \raggedright \footnotesize
    Assuming distance derived from NE2025 model, the Shklovskii effect-corrected spin period derivative can be estimated by $\dot{P}_{\rm shk}=P\mu^2 d/c$ \citep{1970Shklovskii}, which is $\approx$7.8\% of the observed $\dot{P}$.

\label{table:parameters}
\end{table*}
\subsubsection{Eclipses}

\begin{figure*}
    \centering
    \includegraphics[width=0.9\textwidth]{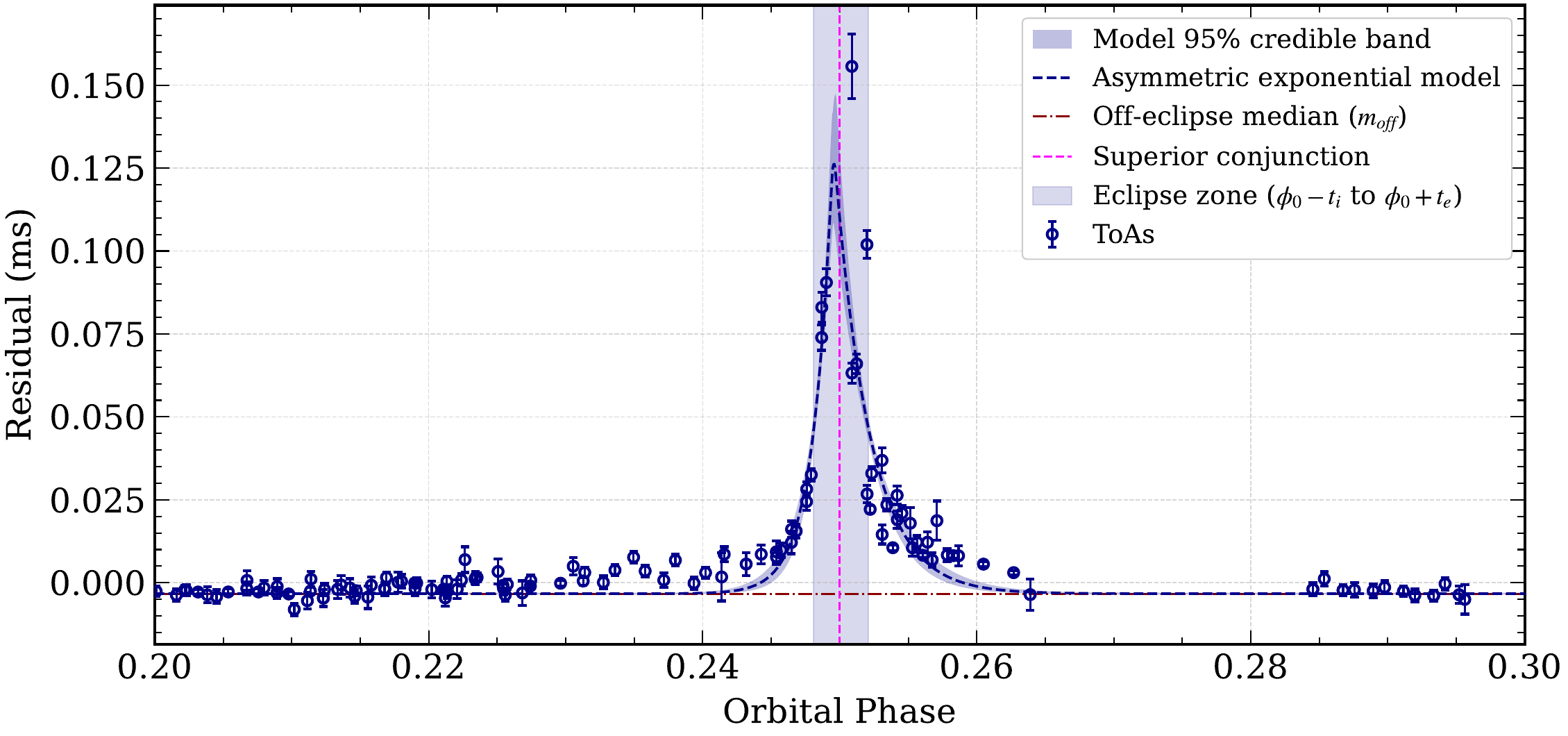}
    \caption{A magnified view of the pre-fit timing residuals for \apsr centered on the superior conjunction. The individual ToAs correspond to an averaged duration of 1.5\,minutes for optimum sampling of the eclipsing zone. The dashed curve and band show the best-fit model and its 95\% credible region. This model is an asymmetric exponential, continuous at its peak phase: $\phi_0$: $m_{off} + A\exp[(\phi-\phi_0)/t_i]$ for $\phi<\phi_0$ (ingress) and $m_{off} + A\exp[-(\phi-\phi_0)/t_e]$ for $\phi\geq\phi_0$ (egress). The baseline $m_{off}$ is the median of the residuals with $|\phi -0.25|<0.1$, weighted by their uncertainties. The best-fit parameters are $\phi_0=0.2495\pm0.0001$, $A=0.135\pm0.009$\,ms, $t_i=(1.43\pm0.14)\times10^{-3}$ and $t_e=(2.57\pm0.22)\times10^{-3}$. The shaded region, from $\phi_0 -t_i$ to  $\phi_0 + t_e$, gives the representative width of the eclipse zone.}
    \label{fig:eclipse_duration}
\end{figure*}

\par
A few of our observations sample the region around the superior conjunction, i.e., the orbital phases ($\phi$) around 0.25. While the ToAs from these observations were excluded in the above-discussed timing of \apsr, these provide useful information about the companion and the environment in the binary system. The ToAs at these eclipsing epochs show a sharp, localized increase in the timing residuals centered at superior conjunction; see Figure~\ref{fig:eclipse_duration}. The exponential rise in the residuals just before the superior conjunction is primarily driven by the excess dispersive delay as the pulsar's signal propagates through the dense intra-binary material. A similar exponential fall-off can be seen in the residuals right after the superior conjunction. Modelling it with an empirical asymmetric exponential, we find that the eclipse is centered at $\phi_0 = 0.2495 \pm 0.0001$ and spans $t_i + t_e = 0.40 \pm 0.03\%$ of the orbit at Band-4. Detailed probes of the frequency-dependent eclipsing behavior using multi-frequency observations obtained from GMRT and the Five-hundred-meter Aperture Spherical radio Telescope (FAST) \citep{2013FAST}, covering the frequency range of 300$-$1500\,MHz, are in progress and will be reported elsewhere (Shrivastava et al. 2027 (in prep.)).

\subsection{Multi-wavelength Counterparts}

\subsubsection{Gamma-ray}\label{sec:gamma_detection}

\begin{figure*}
    \centering
    \includegraphics[width=0.9\textwidth]{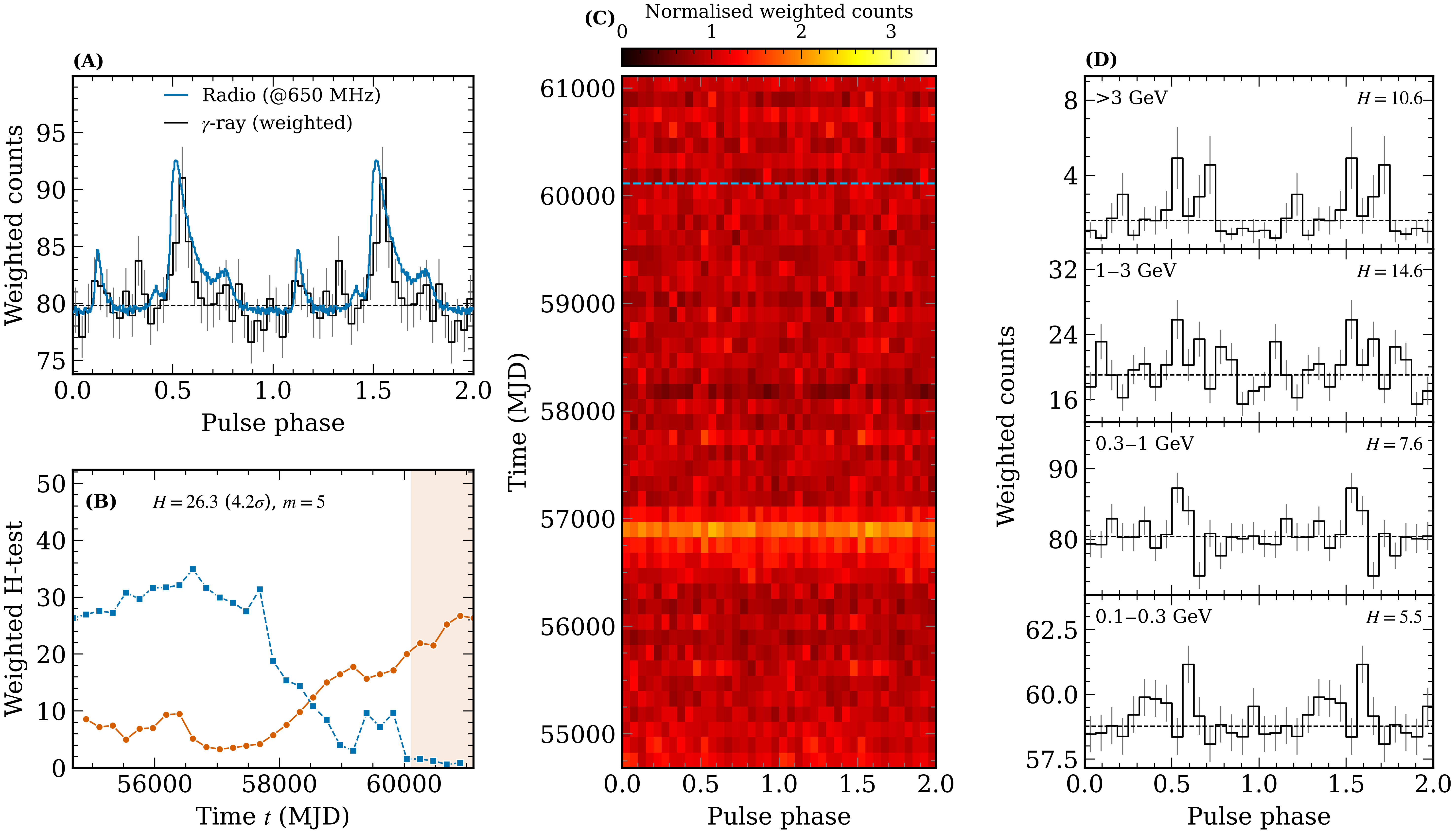}
    \caption{Fermi-LAT pulsations from \apsr, using the simple photon weights of \cite{2019Bruel_weights}. (A) Weighted pulse profile over two rotations, with $1\sigma$ uncertainties; the dashed line is the mean level outside phase 0.4--0.7. A radio profile is overlaid (blue) corresponding to one of the high SNR epochs. (B) Weighted H-test as a function of time, cumulating photons recorded before (orange) and after (blue) time $t$; the shaded region marks the validity range of the radio ephemeris. (C) Weighted photon counts versus phase and time, normalized to their mean; dashed line marks ephemeris validity range. (D) Weighted pulse profiles in four energy bands, with the H-test value for each band.}
    \label{fig:gamma_ray_detection}
\end{figure*}

\par Using the above-derived timing solution on the Fermi Large Area Telescope (LAT) data \citep{Fermi}, we conducted a targeted search for gamma-ray pulsations over the full mission time span. We retrieved LAT photon data with a spacecraft file at 30s resolution, covering MJD from 54683 to 61110. Using \textsc{Fermitools}, we selected with \texttt{gtselect} the events with energies between 0.1 and 1000\,GeV, directions within $3^\circ$ of the radio timing position, and zenith angles below $105^\circ$. Good time intervals were applied with \texttt{gtmktime}, without an additional zenith cut. The final event list contains 411823 photons. Barycentering and pulse phase calculation are done by \textsc{tempo2} using the Fermi plugin \citep{2011Ray}. Each photon was assigned the ``simple weight" of \cite{2019Bruel_weights}, which approximates the probability that it comes from the pulsar using only photon energy and angular separation from the pulsar position.

\par We used the weighted H-test with the maximum number of harmonics, $m=20$, as in the Third Fermi LAT Catalog of Gamma-ray Pulsars \citep[3PC,][]{2023Smith_3PC}. The weighted H-test gives a test statistic of 26.3 with 5 harmonics, corresponding to 4.2$\sigma$; see Figure~\ref{fig:gamma_ray_detection}. For comparison, 3PC adopts $H > 25$ as the detection threshold for simple weight searches with radio ephemeris. Hence, we regard this as evidence for gamma-ray pulsations from \apsr. Furthermore, the gamma-ray and radio profiles seem to be aligned, with the difference ($\delta$) being 0.031$\pm$0.024.

\subsubsection{Optical}

\par Optical detection is essential for exploring the nature of the binary companion. Multi-wavelength characterization of the companion provides a complementary approach to understanding the orbital dynamics and evolutionary history of the system. As the optical emission from the neutron star is typically too faint to be detected, any detected optical flux from the system is heavily dominated by the companion star. We performed a search for optical and infrared counterparts at the pulsar's determined timing position using the ALADIN Lite tool \citep{ALADIN} across photometric data from several archival sky surveys. Despite the depth of the Pan-STARRS survey \citep{chambers2019panstarrs1surveys}, which has a limiting magnitude of approximately 24 in the $g$-band, no counterpart was detected within a 2 arcseconds radius centered on the timing position.

\subsection{Propagation effects} \label{propEffects}
\subsubsection{Interstellar scattering}\label{scattering}
When pulsed radio signals travel through the Interstellar Medium (ISM), they get affected due to plasma dispersion, free-free absorption, and scattering. Scattering is mainly caused by the irregular distribution of electron density in the ISM, giving rise to random variation in the refractive index of the medium, leading to multi-path propagation. The effect of scattering manifests as a temporal broadening of the pulse profile and shows up as an exponential decay of the pulse \citep{1968Natur.218..920S}, with a characteristic timescale known as scatter-broadening time ($\tau$). This scatter-broadening time is strongly dependent on frequency ($\nu$), and scales as $\nu^{-\alpha}$, where $\alpha$ is the power-law index and generally has a value of 4.4 if the ISM irregularities follow a Kolmogorov distribution \citep{1990ARA&A..28..561R}.

To model the scatter-broadening, we first prepared high SNR profiles by integrating the observations across multiple epochs, separately in band-3 and band-4, and divided the individual band data into four sub-bands. We modeled the intrinsic pulse profile as a sum of multiple Gaussian components. The number of components was fixed to six, determined visually by the morphology of the highest SNR sub-band located at 674.2\,MHz in band-4. Components whose fitted amplitudes were found to be below a significance threshold of twice the local off-pulse RMS noise were excluded from the final model. A one-sided exponential decay function was used as the pulse broadening function, and it was convolved with each of the Gaussian components \citep{2001ApJ...562L.157L}. For a Gaussian component with amplitude $A$, center position $\mu$, and a standard deviation $\sigma$, convolved with the above pulse broadening function with a scattering timescale of $\tau$, the scattered intensity $I(x)$ at phase $x$ is given by

\begin{equation}
\begin{split}
    I(x) = A \frac{\sigma}{\tau} \sqrt{\frac{\pi}{2}}\exp\left( \frac{\sigma^2}{2\tau^2} - \frac{x - \mu}{\tau} \right) \times \\ \left[ 1 - \text{erf}\left( \frac{\sigma}{\sqrt{2}\tau} - \frac{x - \mu}{\sqrt{2}\sigma} \right) \right]
\end{split}
\end{equation}

where $\text{erf}(x)$ is the error function.
\par
\begin{figure}
    \centering
    \includegraphics[width=0.45\textwidth]{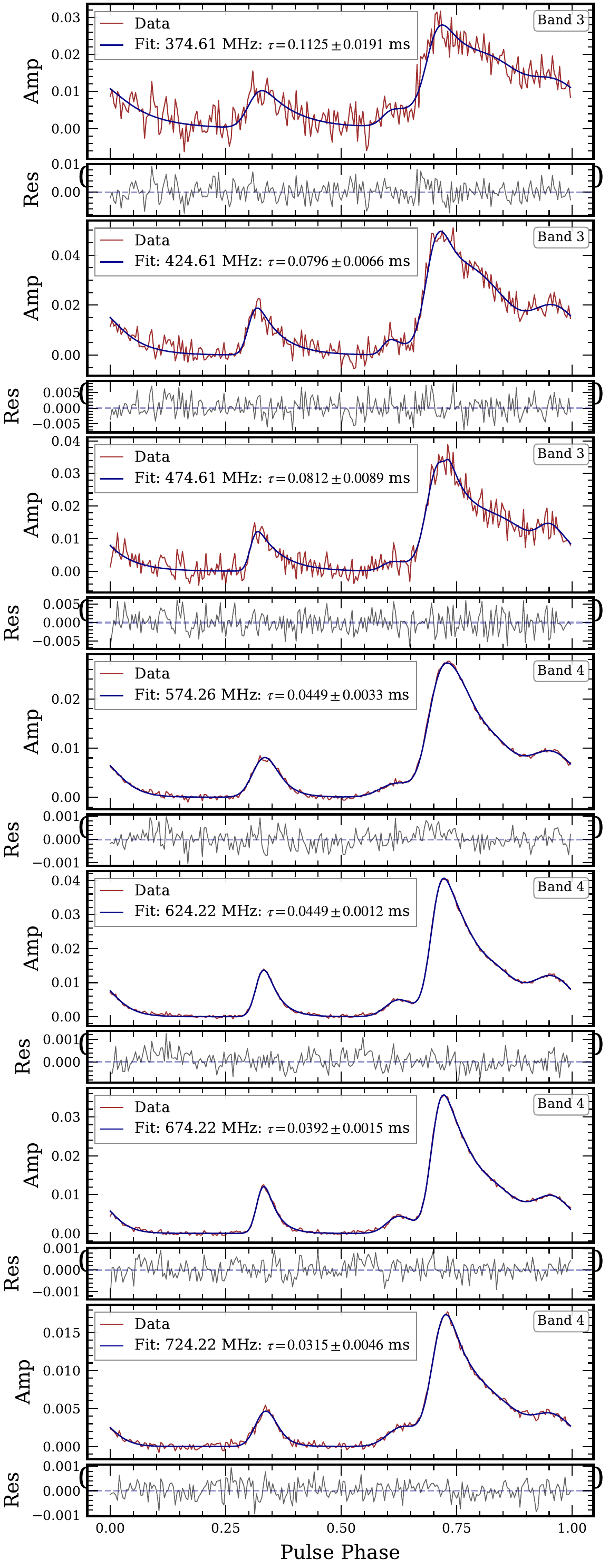}
    \caption{Scatter-broadening fits for average profiles from the individual frequency sub-bands. The set of panels for individual sub-bands show the observed profile in brown color and the best-fit model in blue color in the top panel, with the corresponding residuals plotted in the lower panel.}
    \label{fig:scattering}
\end{figure}

The initial guess for $\tau$ was estimated from the component located at phase $\sim$ 0.32 (see Figure~\ref{fig:scattering}), by measuring the phase distance from the peak to its 1/$e$ intensity drop-off. This particular component was chosen as it is relatively isolated from the complex main pulse profile, providing a clear baseline to reliably quantify the exponential scattering tail. We then obtained the initial parameter estimates using a non-linear least-squares approach using \texttt{curvefit} in the \texttt{scipy} package in Python. 

\par The above estimates were used to initialize a Markov Chain Monte Carlo (MCMC) sampler, implemented via the \texttt{emcee} package \citep{2013emcee}. This Bayesian framework allowed us to robustly explore the multi-dimensional parameter space and prevent ``label-switching" of the peaks by enforcing strict positional priors ($\mu_i < \mu_{i+1}$). The final best-fit parameters were taken as the medians of the posterior distributions, with 1$\sigma$ uncertainties on the two sides marking the 16th and 84th percentiles. The final fitted models as well as the residuals for the individual sub-bands are shown in Figure~\ref{fig:scattering}. The individual fitted components, along with their fitted positions and widths, are shown in the Appendix (Figure~\ref{fig:pos_wid_evo}).

\par
\begin{figure}
    \centering
    \includegraphics[width=0.45\textwidth]{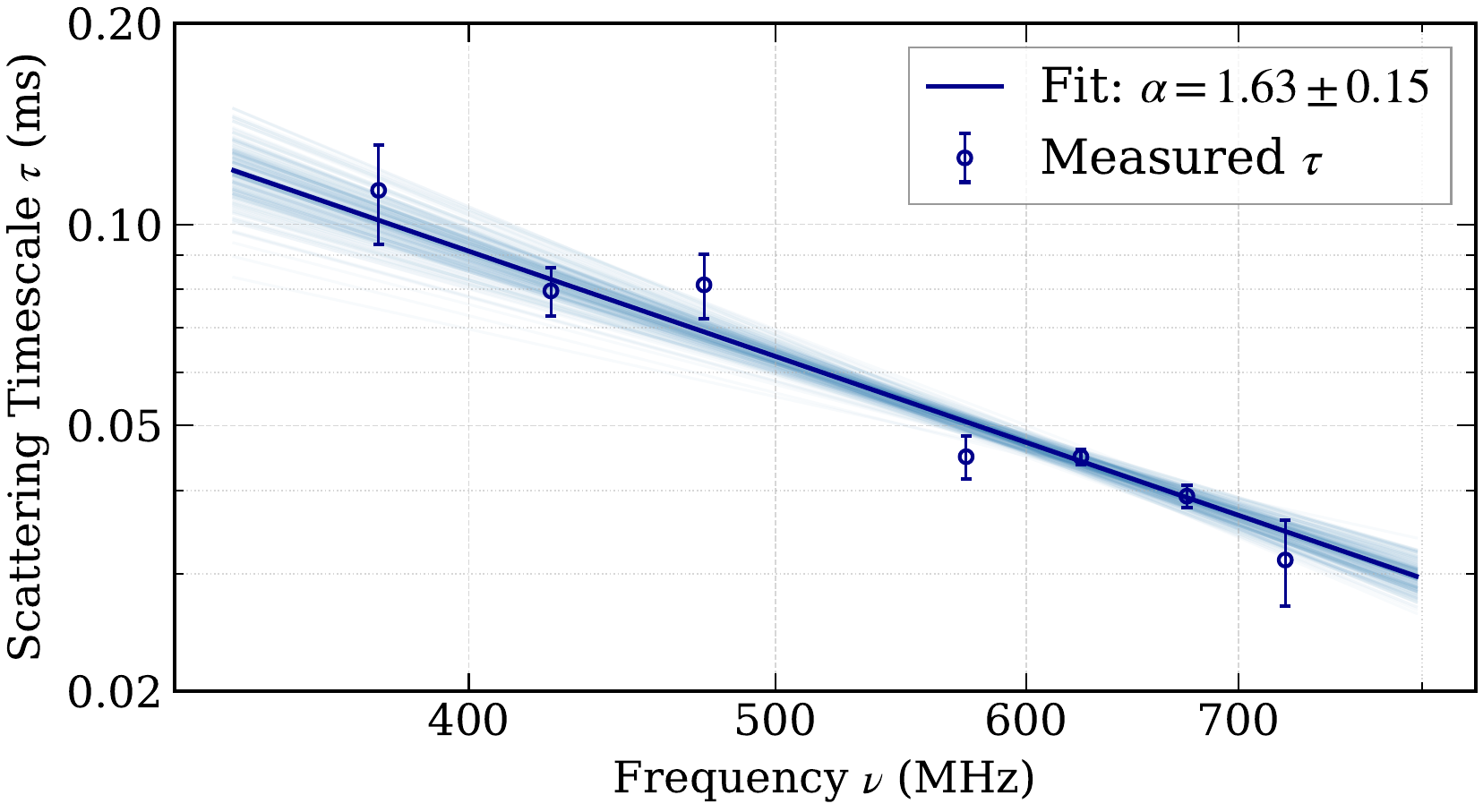}
    \caption{The frequency dependence of the estimated scattering timescale $\tau$. The solid line represents the median MCMC power-law fit, surrounded by a $1\sigma$ confidence envelope drawn directly from the posterior distribution.}
    \label{fig:scat_index}
\end{figure}

To estimate the scattering power-law index, $\alpha$, the frequency dependence of the best-fit $\tau$ was modeled as a power law of the form $\tau(\nu) \propto \nu^{-\alpha}$. We note that the average profile in the lowest-frequency sub-band of band-3 (centered at 324.6\,MHz) had a low SNR and it did not provide a statistically significant measurement of $\tau$. Thus, we excluded this sub-band while modeling the frequency dependence of $\tau$. This power law was fitted directly in the linear space using an MCMC framework. The $\tau$ estimates as a function of frequency as well as the best-fit power law model, corresponding to $\alpha=1.65\pm0.15$, is shown in Figure~\ref{fig:scat_index}.

\subsubsection{Scintillation}

Given our measurements of $\tau$ above, the decorrelation bandwidth due to diffractive scintillation ($\approx 1/2\pi\tau$) at 650\,MHz, i.e., at the band-4 center, is expected to be only a few kHz. Given our typical frequency resolution of nearly 1.5\,MHz, many `scintles' would be averaged within each of our observed frequency channels, and hence, diffractive scintillation is unlikely to be detectable in our observations. 

\begin{figure}
    \centering
    \subfigure{\includegraphics[width=0.47\textwidth]{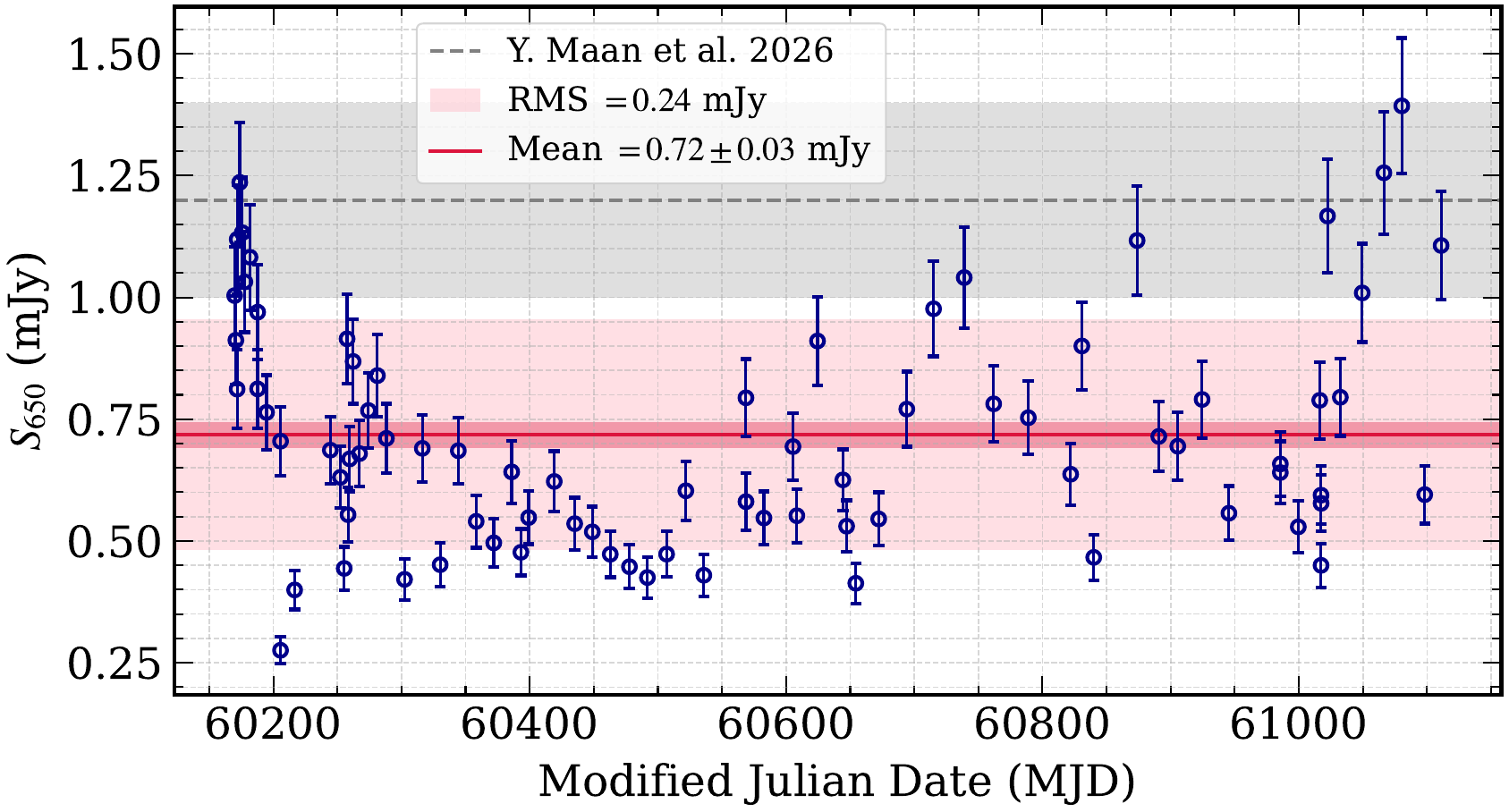}\label{fig:flux-variation}}
    \subfigure{\includegraphics[width=0.45\textwidth]{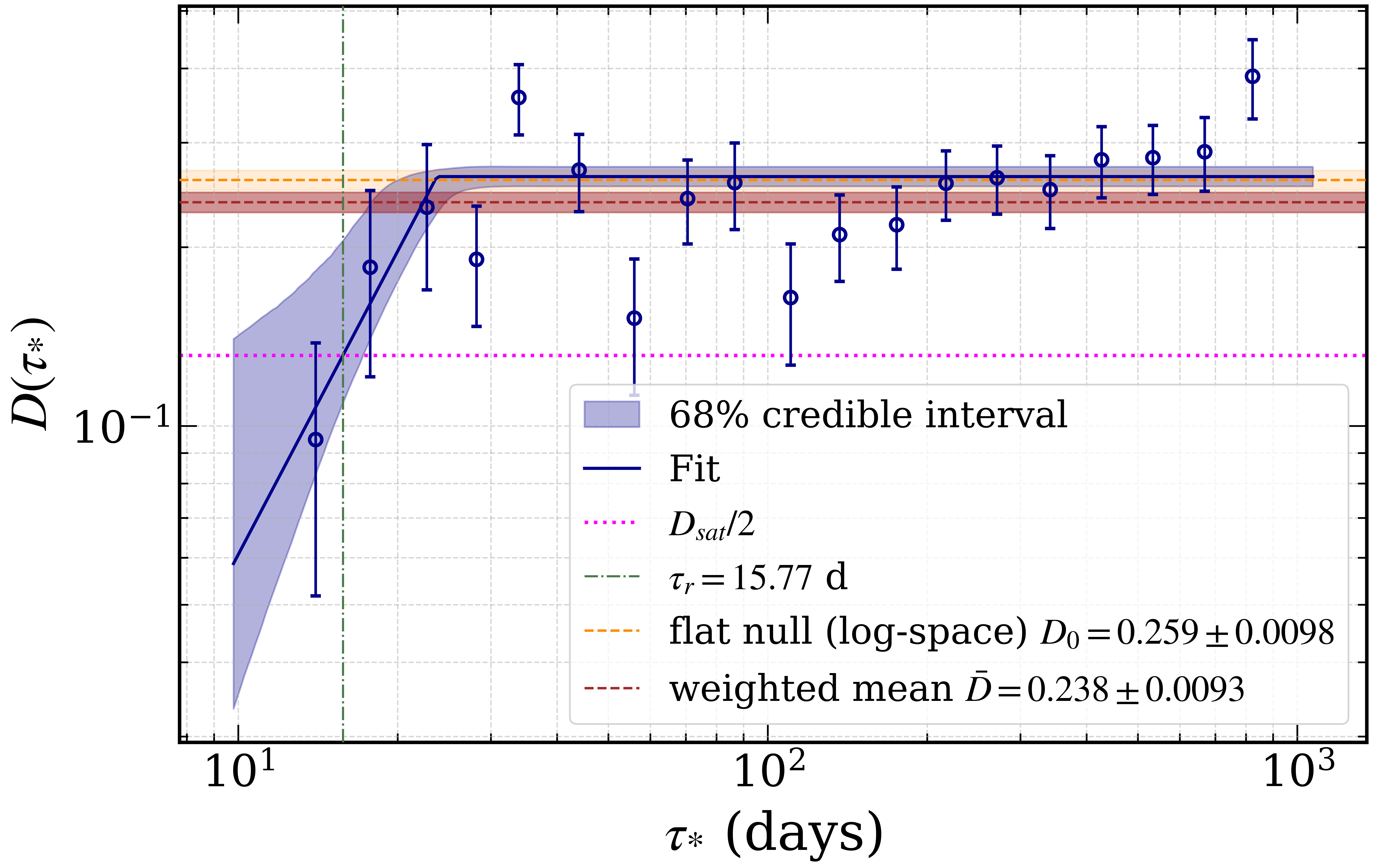}\label{fig:structure_function}}
    \caption{Top: flux density at 650\,MHz ($S_{650}$) calculated using the Radiometer equation with assumed 10\% errors on estimated values. This calculation involves SNR and width estimated using \texttt{pdmp}. The red line denotes the mean value of $S_{650}=0.72 \pm 0.03$\,mJy, the narrow dark-red band shows the uncertainty on the mean, and the wide light-red band shows the epoch-to-epoch rms scatter of 0.24\,mJy. The grey line denotes $S_{650}=1.2\pm0.2$\,mJy as reported by \cite{2026SCOPE}, who note that the flux density is highly variable, consistent with the variability seen in this work. Bottom: the flux density structure function $D(\tau_{*})$ as a function of time lag between observations $\tau_{*}$. Logarithmic binning is used, and error on $D(\tau_{*})$ is propagated as described in equation \ref{eq:err_struc_func}.}
    
\end{figure}

The refractive scintillation primarily manifests itself as amplitude modulations at long timescales, often varying from a few weeks to years. To explore the refractive scintillation effects in our \apsr observations, we estimate the mean flux density at the individual epochs (see Figure \ref{fig:flux-variation}), using the radiometer equation,

\begin{equation}
    S_{\rm mean} = \frac{\mathrm{SNR} \times T_{\rm sys}}{GN\sqrt{n_pt_{\rm int}\Delta f}}\sqrt{\frac{W}{P-W}}
    \label{eq:Flux}
\end{equation}

where $T_{sys}$ is the system temperature, G is the single-antenna gain, N is the number of antennas used in the phased-array observations (typically 20$-$23 for our observations), $n_p$ is the number of polarizations, $t_{int}$ is the integration time, $\Delta f$ is the bandwidth of observation, W is the average profile width, and P is the spin period. 

\par The mean flux density over our 80 epochs is $S_{650}=0.72 \pm 0.03$\,mJy with individual epochs ranging from 0.28 to 1.39\,mJy. This mean is lower than the $S_{650}=1.2\pm0.2$\,mJy reported by \cite{2026SCOPE} which was measured from interferometric images at a few initial epochs. The estimate here is obtained by averaging the pulsed flux density measurements over $\sim$2.6\,years. \cite{2026SCOPE} noted that the flux density varied by a factor of 2.75 between two epochs separated by about a month, and their value lies within the range spanned by our measurements, so the offset can be attributed to sampling a highly variable source. Scaling $S_{650}$ from this work using spectral index $-2.8\pm 0.5$ quoted in \cite{2026SCOPE} gives flux density at 1400\,MHz, $S_{1400} = 84 \pm 30$\,$\mu$Jy, consistent with the 50--200\,$\mu$Jy range established in \cite{2026SCOPE}.

\par To calculate the refractive scintillation timescale, one of the common approaches is to calculate the flux density structure function $D(\tau_{*})$ for a time lag $\tau_{*}$ between each observation epoch, which gives a measure of relative change as a function of time. This is advantageous as the analysis is not affected by non-uniform sampling of the time series. The characteristics of this plot include a noise regime at short time lags, a growing regime at intermediate time lags, which ultimately reaches saturation at longer time lags \citep{1990Stineberg_Condon}. We used the same formalism for calculation as described by \cite{2007You}, and as mentioned in the equations below:

\begin{equation}
\begin{aligned}
    D(\tau_{*}) = \frac{1}{\bar{S}^2N_p}\Bigg[\sum_{i,j}^{N_p} (S_i -S_j)^2 + \sum_{i,j}^{N_p}(e_i^2 +e_j^2)\\
    + 2\sum_{i,j}^{N_p}(S_i-S_j)(e_i -e_j) - 2\sum_{i,j}^{N_p}e_ie_j]\\
\end{aligned}
\label{eq:struc_func}
\end{equation}

\begin{equation}
\begin{aligned}
    \sigma_{D(\tau_{*})}^2 = \frac{1}{\bar{S}^4 N_p^2}\Bigg[\sum_iN_i^2e_i^2 + 4\sum_{i,j}e_i^2(S_i-S_j)^2 +\\ 4\sum_{i,j}^{N_p}e_i^2e_j^2]
\end{aligned}
\label{eq:err_struc_func}
\end{equation}

where $\bar{S}$ is the mean flux density averaged across all observing epochs, $S_i, S_j$ are measured flux densities at individual epochs i and j which are separated by time lag $\tau_{*}$, $e_i, e_j$ are the corresponding measurement uncertainties, $N_p$ is the total number of unique observation pairs that fall into a given time lag bin, $N_i$ is the number of times a specific individual epoch $i$ is used to form pairs within a bin.

For fitting, we used a broken power-law-rise plus saturation (BPL+S) leading to saturation in log-log space, with an initial number of bins of 50. We excluded the points with fractional errors greater than 50$\%$, which resulted in only 19 bins. Further, to check whether the observed structure function shows a genuine rise leading to saturation, we compared the previous model against null model in which $D(\tau_{*})$ is a constant (flat-null) in log-space (limiting case of BPL+S). The comparison of BPL+S with this model gave the difference in Akaike Information Criterion ($\Delta$AIC) and Bayesian Information Criterion ($\Delta$BIC) values of 1.85 and -0.04, which were inconclusive.

\par 
Following \citet{2001Stineberg}, the refractive scintillation timescale ($\tau_r$) is estimated using the value of saturation from the BPL+S model, defined as the point where $D(\tau_{*})$ reaches half its saturation value ($D_{sat}$). The value of $\tau_r$ is estimated around 15.77$^{+2.60}_{-4.42}$ days (see figure \ref{fig:structure_function}). The uncertainties are derived from 5000 Monte Carlo samples drawn from the BPL+S fit covariance matrix, giving all plausible values of $\tau_r$, with the 68$\%$ interval referenced to the best-fit $\tau_r$. The reduced $\chi^2$ value for this fit is 1.7. 

\section{Discussion}\label{sec:discussion}

We have presented a detailed timing study of the recently discovered binary millisecond pulsar \apsr, its multi-wavelength counterpart search, along with a study of the ISM toward this source. Using timing parameters, we have also derived various other quantities, as discussed below.

\par
\begin{figure}
    \centering
    \includegraphics[width=0.47\textwidth]{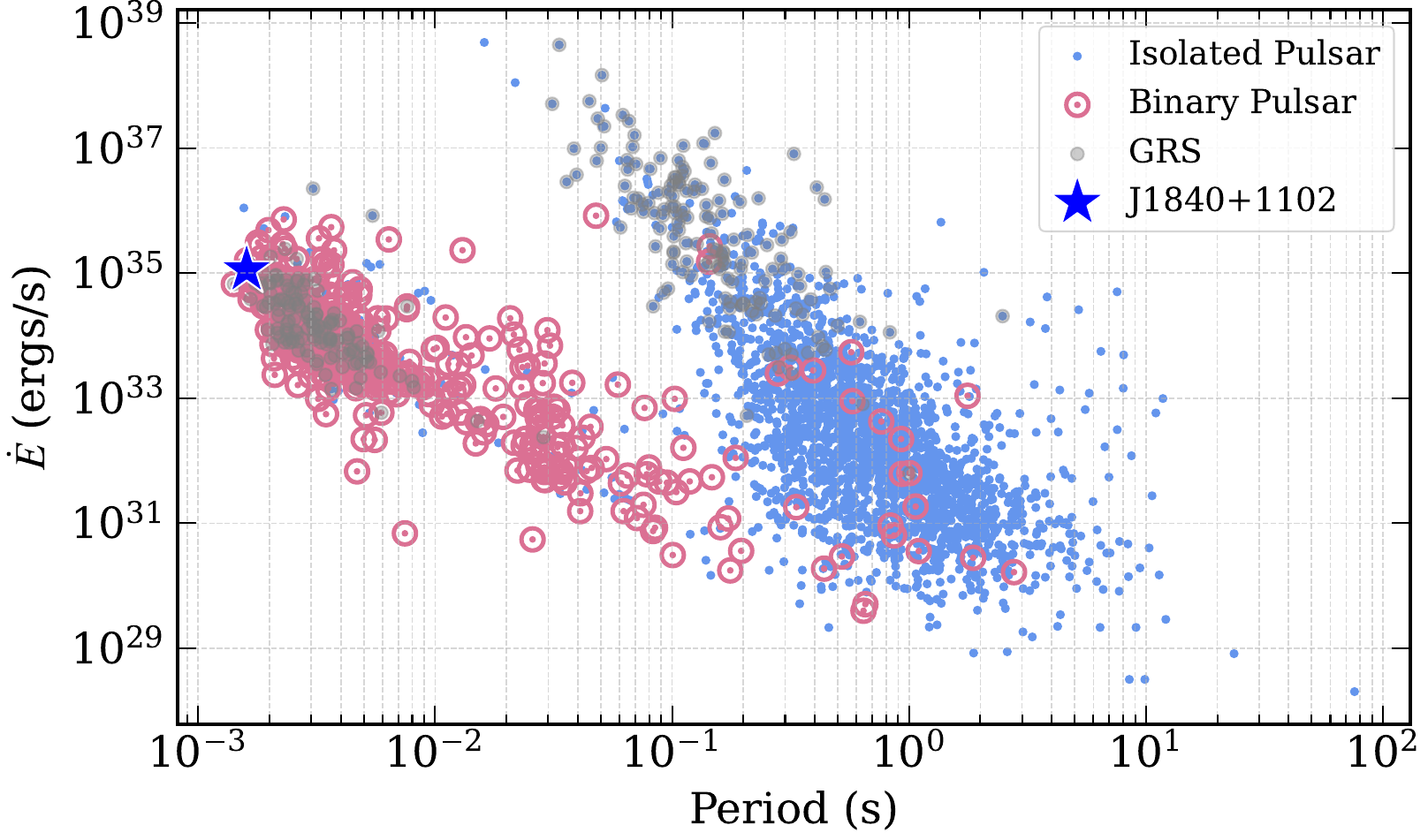}
    \caption{Spin-down luminosity ($\dot{E}$) vs. the spin period (P) for the known pulsars. The pulsars from which pulsed gamma-ray emission has been detected are marked as gamma-ray sources (GRS); \apsr is marked as a star.}
    \label{fig:Edot_P0}
\end{figure}

 The estimated spin-down luminosity ($\dot{E}$), assuming a standard neutron star moment of inertia ($\simeq 10^{45}$\,g\,cm$^2$), is $1.1\times10^{35}$\,erg\,s$^{-1}$. As evident from Figure~\ref{fig:Edot_P0}, \apsr's $\dot{E}$ is at the higher end and lies in the cluster of pulsars known to emit in gamma-rays. This fueled the idea of gamma-ray search, and using the full time span of Fermi-LAT data, we discovered pulsed gamma-ray emission, as discussed in \ref{sec:gamma_detection}. We compare its gamma-ray properties with the 3PC. At $P=1.60$\,ms, \apsr is among the fastest-spinning gamma-ray MSPs. Its $\dot{E}$ is at the 94th percentile of the 144 3PC MSPs. Radio and gamma-rays arrive at the same phase ($\delta=0.031 \pm 0.024$). Only around 12 of the 103 MSPs with a measured $\delta$ have $|\delta|<0.05$, so \apsr is among the $\sim12\%$ having the most aligned radio and gamma-ray emission. Aligned radio and gamma-ray profiles are seen mainly in energetic MSPs, such as PSR~J1939+2134 (B1937+21) and PSR~B1821$-$24 \citep{2012Guillemot, 2013Johnson}. In such pulsars, the radio and gamma-ray emission are expected to originate from co-located regions, potentially at high altitudes close to the light cylinder \citep[see, e.g.,][]{2012Venter}. Although not very significant, visual evidence for a weaker gamma-ray component aligned with a secondary, isolated radio component ($\Delta \approx 0.5$) exists, supporting the possibility of a double-peak gamma-ray profile. Using the spectral parameters of the associated source  J1840.2+1104 ($0.053^\circ$) listed in the Fermi-LAT 16-Year source list \citep[][FL16Y]{2026FL16Y}, we also calculated the SED peak energy ($E_p\approx1.9$\,GeV) and curvature ($d_P\approx0.46$) and they also agree with the 3PC $E_p-\dot{E}$ and $d_p-\dot{E}$ trends. Its luminosity ($\approx 10^{33}$)\,ergs/s and efficiency are consistent with the MSPs of similar $\dot{E}$ for DM distance estimated using YMW16 with $f_\Omega =1$, corresponding to an efficiency of $\eta = 7\%$.

\par 
Notably, the two MSPs with aligned radio and gamma-ray profiles mentioned above, PSR~B1937+21 and PSR~B1821$-$24, are also the best-known MSP sources of giant pulses, and have similarly high magnetic field strength at the light cylinder radius, $B_{LC}$ \citep{1996Cognard_GPs, crawford}. While this makes PSR J1840+1102 a promising candidate for such phenomena, a systematic search for giant pulses in many of our initial observations did not result in any bright narrow pulses. We note that our sampling time ($\simeq$5.12\,$\mu$s) is comparable to the scatter-broadening, and is unlikely to be the reason for non-detection of giant pulses at our observing frequency. 


\subsection{Characterization of the binary system}

\par The critical factor in understanding the formation of a recycled pulsar is its spin period, which directly reflects the amount of mass the neutron star has accreted. However, the rate, efficiency, and duration of this mass transfer phase are fundamentally dictated by the nature of the donor star.

\par 
Characterizing the companion is essential for constructing the system's mass transfer history. Binary pulsar systems can host a variety of companions such as a white dwarf (WD), main-sequence (MS) star, planet, or another neutron star (NS), depending on the evolutionary track followed by the progenitor stellar pair \citep{2005Lorimer_review}. Binary parameters such as the orbital period ($P_B$), eccentricity ($e$), mass of the companion ($M_c$) and the inclination angle ($i$) of the orbit from the sky plane are some essential ingredients to study the evolutionary timeline of the system. Most accurate estimates for $M_c$ are generally determined from Shapiro delay \citep{1964Shapiro} measurements. However, that is typically possible only for nearly edge-on systems and is achievable after long-term, highly precise pulsar timing \citep{2020Cromartie}. Hence, the best way to constrain the companion mass and type is to use orbital parameters in conjunction with the inclination angle derived from the optical data. If no optical counterpart is detected, the mass function can provide the median companion mass for $i=60^\circ$, as randomly aligned orbits on the sky will imply a uniform distribution in $cos(i)$. Assuming a canonical neutron star mass of $1.35 M_\odot$, the estimated maximum, median and minimum M$_c$ for the \apsr system are approximately $0.41, 0.19\text{ and }0.16 M_\odot$, corresponding to orbital inclination angles of $26^\circ, 60^\circ\text{ and }90^\circ$, respectively.

\par In addition to the mass estimate, the short spin period of \apsr confirms its evolution via the standard low-mass X-ray binary (LMXB) channel, producing a fully recycled MSP. Below, we evaluate the viability of various companion types, drawing on the evolutionary frameworks discussed by \cite{2005van} and \cite{2025Wang}. To assess the detectability of these potential companions, we first estimate their expected apparent magnitudes. using the pulsar distance derived from the YMW16 \citep{2017YMW16} and NE2025 \citep{NE2025} electron density models (5.6 and 5.5 kpc, respectively), the distance modulus ($5\log_{10}(d) - 5$) contributes around 13.7$-$13.8 magnitudes. Furthermore, as the pulsar resides at a relatively low Galactic latitude ($l,b = 41.537^\circ, 7.485^\circ$), interstellar extinction will significantly obscure optical emission. Assuming a conservative average visual interstellar extinction of $A_V \approx 1.0$ mag kpc$^{-1}$, the total extinction along this line of sight adds an additional 5.5 to 5.7 magnitudes of dimming.

\begin{itemize}
    \item Ultra-light (UL) companion: UL companions are generally formed via a particular case of Roche-lobe overflow in LMXBs, where mass transfer initiates while the donor star is still on the main sequence \citep{2023Tauris_book}. The orbit subsequently becomes compact due to angular momentum loss driven by magnetic braking, leaving a remnant with a median mass of $\lesssim 0.08 M_\odot$ \citep{Manchester_2005, 2012Tauris}. Given our derived minimum companion mass constraint of $0.16\,M_\odot$, an ultra-light companion for PSR J1840+1102 is firmly ruled out.

    \item CO or ONeMg WDs: Intermediate-mass X-ray binaries (IMXBs) typically evolve to leave massive white dwarf companions, specifically CO-WDs with typical $M_c \gtrsim 0.5 M_\odot$ or ONeMg-WDs with typical $M_c \gtrsim 1.05 M_\odot$ \citep{2022Camisassa_CO/ONeMg}. Pulsars in such systems exhibit a broad range of spin periods (2.9 ms $< P < 1.1$ s) and often possess orbital periods clustering around 10 days. Given the above-discussed companion mass constraints for the \apsr system, hosting a CO or ONeMg WD companion would require a highly improbable, nearly face-on orbital inclination ($i \lesssim 15^\circ$ for a CO-WD, and $i \lesssim 8^\circ$ for an ONeMg-WD). The presence of eclipses observed from the \apsr system makes such low orbital inclinations further unlikely. Therefore, a massive white dwarf companion like CO or ONeMg WD is statistically disfavored by the orbital geometry.

    \item Neutron Star: The possibility of a double neutron star (DNS) system is definitively ruled out for \apsr system on two fronts. First, the typical mass of a neutron star ($\sim 1.35 M_\odot$) far exceeds our maximum companion mass constraint, and it would require an implausibly low orbital inclination. Second, the formation of a subsequent neutron star generally involves an asymmetric supernova explosion that imparts a massive kick to the system. This violent event induces high orbital eccentricity, directly contradicting the highly circular orbit ($e \simeq 1.84 \times 10^{-6}$) observed for \apsr.

    \item Main Sequence (MS) Star: While pulsars with wide orbits ($P_B > 50$ days) and massive companions ($M_c > 0.5 M_\odot$) typically host MS stars \citep{2012Tauris}, MS companions are also frequently found in tight binaries ($P_B \lesssim 1$ day) with low-mass companions ($0.1 < M_c < 0.4 M_\odot$), classically categorized as redback (RB) spider systems. With $P_B \simeq 1.68$ days, \apsr lies beyond the orbital period range typical of RBs. In RBs, the Roche-Lobe-filling companion and its ablated material produce long eclipses, often covering a large fraction of the orbit. In contrast, the short eclipses of \apsr imply an eclipsing region far smaller than a companion's Roche Lobe, arguing against a RB configuration. 
    
    For \apsr, an MS companion within our derived mass range would be classified as an M-type red dwarf, \citep{2010Bochanski_Mdwarf}. These cool, low-luminosity stars are intrinsically faint. Even at the upper mass limit of $0.41 M_\odot$, an M-dwarf possesses an absolute visual magnitude of only $M_V \approx 10$. Applying our distance and extinction estimates, the brightest possible apparent magnitude for an M-dwarf in this system would be $m_V \approx 29$, falling significantly below the Pan-STARRS detection threshold ($m_V \approx 24$). Hence, a low-mass main-sequence companion remains a physically viable, albeit currently unobservable, possibility. 
    
    \item Helium~WD: The majority of fully recycled pulsars in the Galactic field, once accretion has ceased, host low-mass He~WD companions \citep{2006csxs.book..623T}. With an orbital period of $\simeq 1.68$ days and a derived minimum companion mass of $\simeq0.16\,M_\odot$, \apsr is in agreement with the theoretical orbital period--mass ($P_B - M_c$) relation for He~WD companions formed through stable, long-term mass transfer \citep{1999Tauris&Savonije, 2023Tauris_book}. In this scenario, during the LMXB phase, the accretion of material and angular momentum from the Roche-lobe-filling secondary star would have successfully spun up the neutron star to its current 1.60\,ms spin period. Furthermore, prolonged tidal interactions during this evolutionary phase could have effectively damped the orbital ellipticity, fully accounting for the observed highly circular orbit. The typical He~WD companions possess absolute visual magnitudes in the range of $M_V \approx 12$ to $13.5$ \citep{2005van}. The potential He~WD companion, placed at the distance of \apsr ($\approx5.5\,kpc$), would have an apparent magnitude of $m_V \simeq 31-32.5$. Given that this apparent magnitude falls several magnitudes below the Pan-STARRS detection threshold, non-detection of an optical counterpart is not unexpected.
\end{itemize} 

\begin{figure*}
    \centering
    \includegraphics[width=0.47\textwidth]{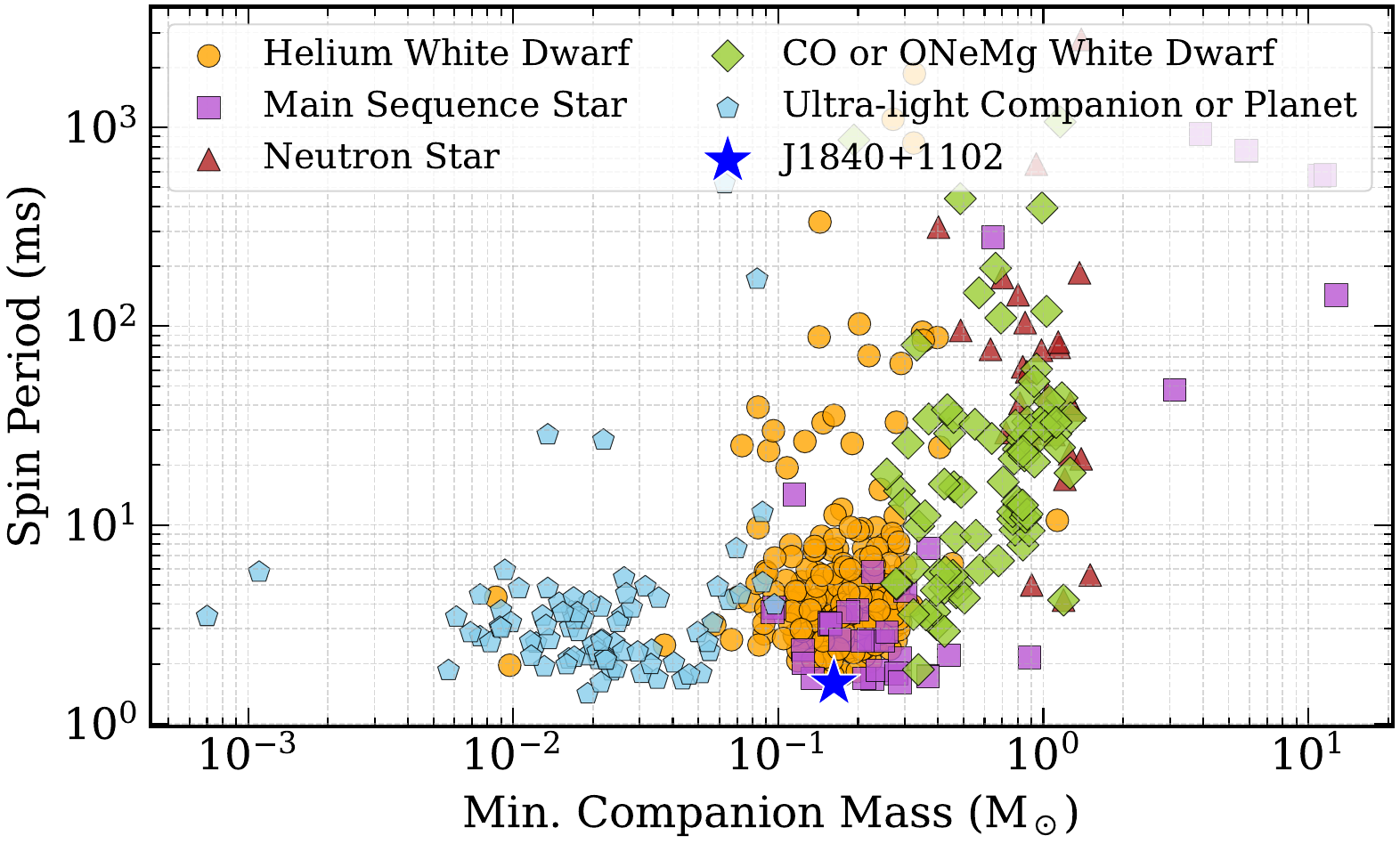}
    \includegraphics[width=0.47\textwidth]{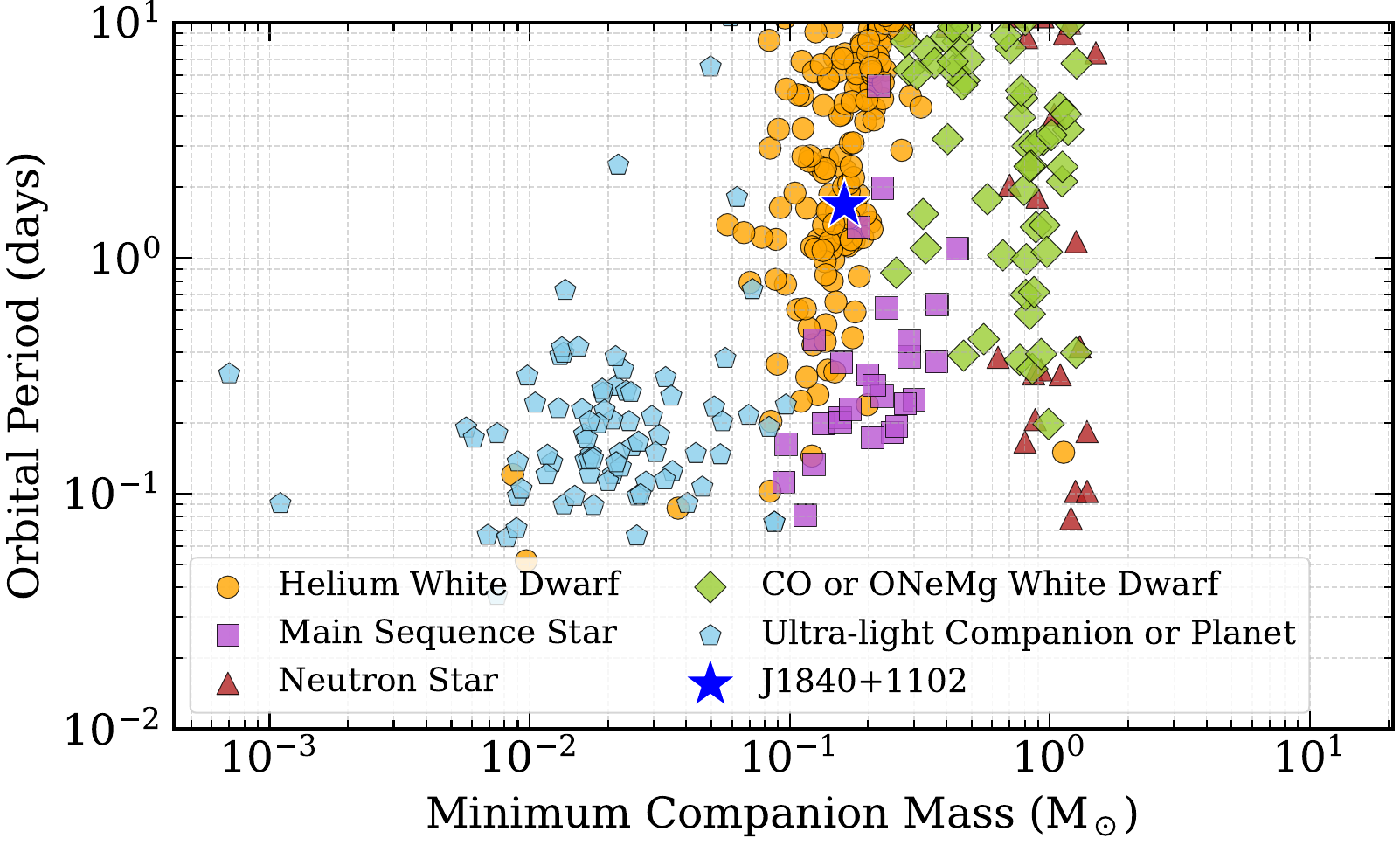}
    \caption{Various pulsar binary systems with known companions in the spin period vs minimum companion mass (left) and orbital period vs minimum companion mass (right) phase spaces. \apsr is shown by a star in both panels.}
    \label{fig:mass_relation}
\end{figure*}

\par We also utilized the ATNF catalogue\footnote{\url{https://www.atnf.csiro.au/research/pulsar/psrcat}} \citep[Version 2.8.0,][]{Manchester_2005} for a population analysis in the $P-M_{c,min}$ and $P_B-M_{c,min}$ phase spaces, where $M_{c,min}$ is the minimum companion mass. As shown in Figure~\ref{fig:mass_relation}, \apsr falls within a region commonly populated by He~WD systems.

\par The extremely short eclipse duration ($t_i + t_e \approx 0.40 \%$ of the orbital period at Band-4) presents an interesting geometric constraint. The separation between the pulsar and its companion is given by $a = a_p(1 + M_p/M_c) \approx 6.9\,R_\odot$ for $i = 90^\circ$ and $M_c = 0.16\,M_\odot$, with an assumed pulsar mass 1.35$M_\odot$. The eclipse then corresponds to a chord of length $\approx 0.17R_\odot$, i.e.\ an characteristic eclipsing region of radius $\approx 0.09R_\odot$. This is more than an order of magnitude smaller than the companion's Roche Lobe radius, $R_L \approx 1.5 R_\odot$ \citep{1983Eggleton} and $\sim 4$ times larger than typical low-mass He-WD ($ \sim 0.02 R_\odot$). An occultation by the WD itself would last only $\sim0.09\%$ of the orbit and would require an edge-on view. Since $t_i +t_e$ traces only the dense core of the eclipsing medium, these sizes are effectively lower limits, implying that the eclipse is caused not by the companion's body but by material extending well beyond its surface.

\par The presence of such material is energetically plausible. We compared $\log_{10}(\dot{E}/a^2)$, a proxy for the spin-down flux incident on the companion, for \apsr with that of black widow (BW), redback (RB) spider systems and WD binaries; refer to Appendix~\ref{irradiation} and Figure~\ref{fig:BW_RB_WD}. The value for \apsr lies at the lower end of the RB distribution and the upper end of the WD distribution, indicating that the pulsar wind could drive substantial ablation. However, the short eclipse argues against a Roche-Lobe-filling companion. Instead, the relatively high incident flux can provide enough energy for any plasma surrounding the He WD companion. 

\par This interpretation is supported by the frequency dependence of the eclipses, which at first sight means the existence of a plasma around a He WD. But this seems surprising too; a high surface gravity prevents a conventional stellar wind. Still, the eclipses may occur like in the double pulsar system PSR J0737-3039A/B, where pulsar A is eclipsed by the magnetosphere of its compact neutron-star companion, showing chromatic behavior \citep{2004Kaspi_doublepulsareclipse}. By analogy, the eclipsing material in \apsr could originate from (i) mass loss from the He WD, driven by relatively high incident spin-down flux; (ii) plasma confined in a WD magnetosphere; or (iii) an interaction region between the pulsar wind and WD. This gives an interesting opportunity to explore MSP-WD interaction. Discriminating between them will require in-depth study of eclipsing material and mechanisms over broad frequency ranges which will be subject of a future work. Furthermore, mass estimates through continued precision timing and deep optical observations will help in robustly determining the type of \apsr's companion.

\subsection{Propagation effects and profile evolution}\label{disc_propeffects}

Our measurements discussed in Section~\ref{propEffects} suggest the scatter-broadening ($\tau$) to be in the ranges of 80$-$113\,$\mu$s and 30$-$45\,$\mu$s at the band 3 and band 4 frequencies, respectively. The $\tau$ measurements increase monotonically as the observing frequency decreases, with the scattering index, $\alpha$, estimated to be $1.65 \pm 0.15$. The scattering index derived for various radio pulsars typically has a value around 4.0 with a standard deviation of about 0.6 \citep{2021MNRAS.504.1115O}. It is predicted assuming a single screen situated midway to the pulsar, which is infinitely extended transversely and thin along the LOS \citep{1972Williamson}. Our measured value of $\alpha = 1.65 \pm 0.15$ is significantly flatter. The lower scattering index can be explained using anomalous scattering mechanisms and geometries, such as truncated screens \citep{2001cordesandlazio} and anisotropic scattering mechanisms \citep{2001Stineberg}. Many studies have also investigated this scaling relation and have found $\alpha$ to be much lower than 4.4 for several pulsars \citep{2001ApJ...562L.157L, 1997Ramchandran, 2004Bhat, 2013Lewandowski}, also stating dependence of calculated $\alpha$ on the choice of PBF for a given LOS. Hence, a mere departure from the traditional $\alpha$ value need not necessarily imply an anomaly for the scattering along that LOS.

\par As presented in Section~\ref{sec:data}, the refractive scintillation time scale, $\tau_r$, is estimated to be 15.77$^{+2.60}_{-4.42}$ days.  However, the slope of the rising regime in the BPL+S model is loosely constrained (= 1.697 $\pm$ 1.049), leading to loose constraints on $\tau_r$ too. Also, the inconclusive result from comparison with the flat-null model indicates that the current data do not provide preference for a genuine rise leading to saturation over a constant $D(\tau)$, and hence the resulting $\tau_r$ should be interpreted with caution. A robust estimate can be made with more observations with shorter time lags in the future.

\par The intrinsic frequency evolution of the pulse profile can be judged by tracking the centroid phase positions ($\mu$) and widths (obtained from fitted $\sigma$) of the individual fitted Gaussian components across the observed sub-bands. As described in Appendix~\ref{components} and apparent from Figure~\ref{fig:pos_wid_evo}, the phase positions of the components remain highly stable across the entire frequency range. The fitted widths exhibit significant variability; however, we note that the effect of covariance between the component width and the scatter broadening could easily explain these variations. The amount of covariance depends on a number of parameters, including the observed SNR, and hence varies slightly from one sub-band to the other. We note that the effect of this covariance on the fitted positions would be much less pronounced.

\par Overall, the intrinsic pulse-components and their widths appear to be largely stable across the observed frequency range of 300 to 750\,MHz. These minimal variations indicate that the radius-to-frequency mapping \citep{1975RFM} is negligible for \apsr. This is not surprising as \apsr, like many other MSPs, has a relatively small magnetosphere with a light cylinder radius of only around 76\,km, and the radio emission across the observed 300$-$750\,MHz range would originate from nearly co-located regions in the pulsar magnetosphere. The lack of profile width variations might also be expected if the radio emission originates at high altitudes, as indicated by the radio and gamma-ray profile alignment.

\subsection{Suitability for PTA experiments}

The achieved RMS of $\sim$2.15\,$\mu$s for \apsr is comparable to, and in some cases even better than, the timing residuals of pulsars currently monitored in the PTA experiments. However, a low post-fit residual RMS over a few years' baseline does not automatically guarantee a pulsar's viability for PTA experiments. Additionally, excluding the eclipse duration, which is only around 0.40\% of the orbit, we are yet to see any evidence for orbital variations widespread over the orbit. Glitches also have not been observed from \apsr so far. While a detailed timing noise analysis is yet to be done, generally well-behaved residuals and the absence of measurable $P_B$ variations over the baseline of 2.7\, years point towards the \apsr's potential for inclusion in the PTA experiments.

\par
\apsr's relatively high DM could make it inherently more susceptible to turbulent interstellar medium (ISM) effects. While the changes in DM can be measured and corrected to a high precision using adequately large bandwidth observations, any changes in scatter-broadening are not trivial to account for in pulsar timing. Long-term follow-up and timing of \apsr will help in quantifying the propagation effects as well as potentially measure the rate of change of $P_B$ and inform better about the suitability of this pulsar for PTA experiments.

\section{Conclusion}\label{sec:conclusion}
We have presented a comprehensive timing and multi-wavelength study of the eclipsing MSP J1840+1102, recently discovered in the SCOPE survey using GMRT. Our primary conclusions are as follows.

\begin{enumerate}
    \item Our analysis of data over a 2.7-year baseline yielded a robust phase-connected timing solution with a post-fit RMS residuals of 2.1\,$\mu$s. By folding 17\,years of Fermi-LAT data with the radio ephemeris, we discovered pulsed gamma-ray emission from \apsr at a 4.2$\sigma$ level. The gamma-ray peak is aligned with the radio profile ($\delta=0.031\pm0.024$), a property shared mainly by the most energetic MSPs and consistent with high $\dot{E}$ of \apsr.    
    \item The 1.68-day orbit is highly circular, and the mass function implies a minimum mass of the companion to be 0.16$M_\odot$. We interpret the companion to be a He~WD formed through stable mass transfer in a LMXB, which explains the spin period, orbital period, as well as the circular orbit. The extremely short eclipses around the superior conjunction (only about 0.4\% of the orbit at 650\,MHz) might suggest a nearly edge-on view of the system. The physical extent of the eclipsing region suggests material around the companion and offers an opportunity to directly study the interaction between He WD and \apsr.
    \item Our measured scatter broadening towards \apsr is $\sim$0.04 ms (at 650\,MHz) and the scattering index ($\alpha$) of $1.65\pm0.15$ is much lower than the Kolmogorov expectation. We also measured a tentative refractive scintillation timescale of $\sim16$\,days, though more high-cadence observations would be needed to confirm this measurement.                  
 \end{enumerate}

Given its well-behaved low timing residuals, and the absence of glitches and measurable orbital variation over the current baseline, \apsr can potentially be added to the ensemble of pulsars used in the PTA experiments. Finally, continued timing and multi-frequency observations of the eclipses will help in constraining various relativistic effects and study the material around the He~WD companion.

\begin{acknowledgments}

A.S. and Y.M. are grateful to David A. Smith and Matthew Kerr for sharing their time and expertise to help with the gamma-ray photon folding analysis, and to David A. Smith for providing useful comments on the manuscript. A.S. would like to thank Ankita Ghosh, Sangita Kumari, and Shalini Sengupta for helping with learning timing during the initial phase of this project. A.S., Y.M., M.A.K. and B.L. acknowledge the support of the Department of Atomic Energy, Government of India, under the project title “Next Generation Instrumentation for Radio Astronomy”, UID number RTI4017. J.v.L. acknowledges support from the project CORTEX with file number NWA.1160.18.316 of the research programme NWA-ORC which is (partly) financed by the Dutch Research Council (NWO) under the grant https://doi.org/10.61686/CJLAY91420. 
The generation of various algorithms used and learning new software packages were supported by Claude Opus 5.5. The GMRT is run by the National Centre for Radio Astrophysics of the Tata Institute of Fundamental Research, India. We acknowledge the support of GMRT telescope staff for observations. 
\end{acknowledgments}

\appendix

\section{Irradiation of the companion}\label{irradiation}

\begin{figure*}
    \centering
    \includegraphics[width=0.4\linewidth]{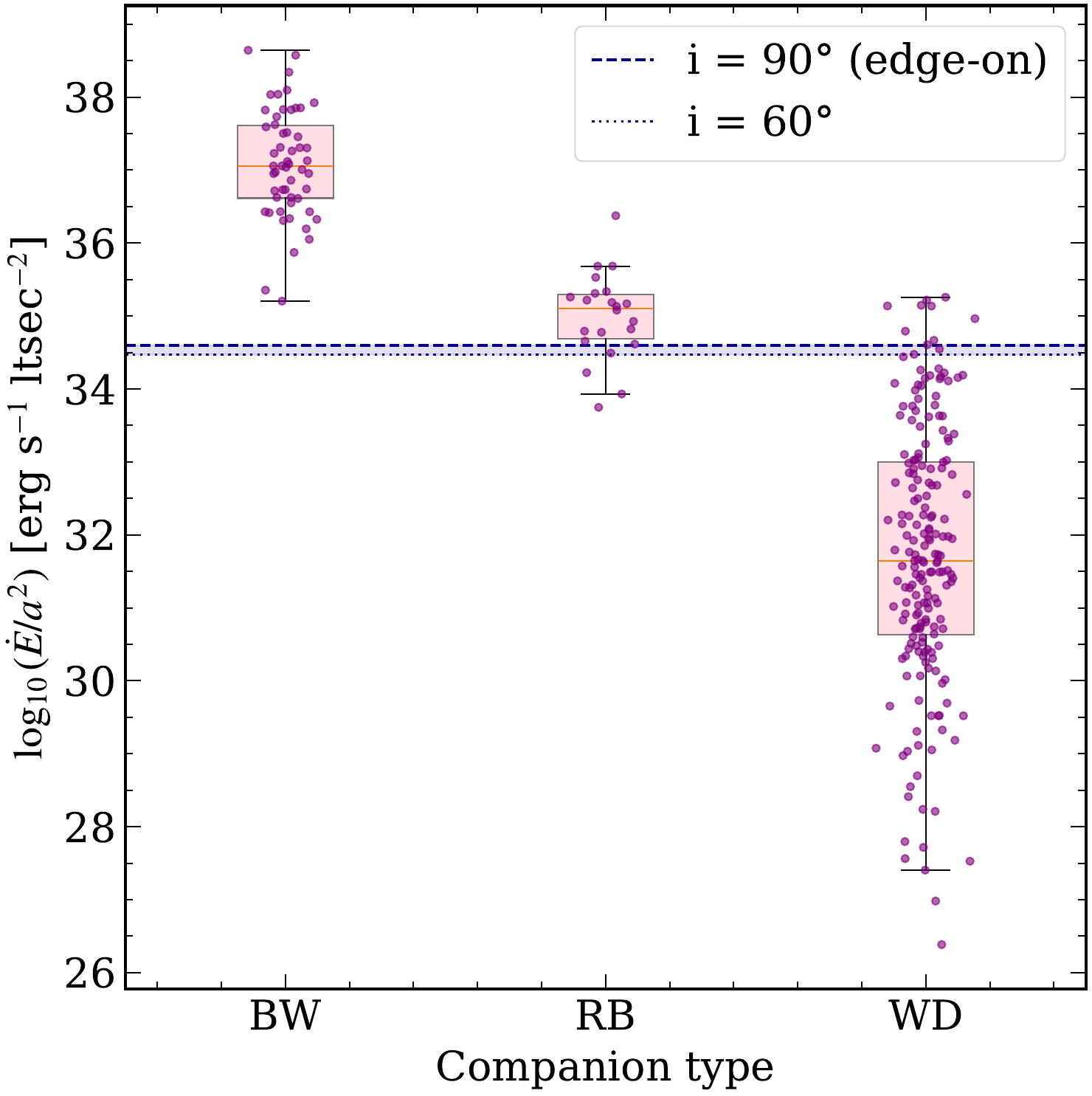}
    \caption{Spin-down luminosity normalized by the square of the pulsar's orbital semi-major axis, $\log_{10}(\dot{E}/a_p^2)$, for binary MSPs with 54 BW, 22 RB and 196 WD taken from the ATNF\footnote{\url{https://www.atnf.csiro.au/research/pulsar/psrcat}} \citep[Version 2.8.0,][]{Manchester_2005} catalogue \citep[Version 2.8.0,][]{Manchester_2005} and SpiderCat\footnote{\url{https://astro.phys.ntnu.no/SpiderCAT}} \citep{2025SpiderCat}. For each population, the box spans the 25th to 75th percentiles, the orange line marks the median, and the whiskers extend to 1.5 times the interquartile range. The horizontal lines show the values for \apsr for $i=90^\circ \text{and } 60^\circ$.}
    \label{fig:BW_RB_WD}
\end{figure*}

To check whether \apsr can supply enough energy to drive a redback-like outflow, we compare its irradiation proxy, $\dot{E}/a_p^2$ (where $a_p=A_1/sini$ is the semi-major axis of the pulsar orbit), with those of BW ($N=54$), RB ($N=22$) and WD ($N=196$) binary pulsars from ATNF\footnote{\url{https://www.atnf.csiro.au/research/pulsar/psrcat}} \citep[Version 2.8.0,][]{Manchester_2005} and SpiderCat\footnote{\url{https://astro.phys.ntnu.no/SpiderCAT}} \citep{2025SpiderCat} Catalogues; see Figure~\ref{fig:BW_RB_WD}. The values for \apsr at different inclination angles fall just below the interquartile range of the RBs ($\sim0.9\sigma$ below its median), but in the upper tail of the WD population ($\sim1.6\sigma$ above its median), and well below typical BWs. \apsr is irradiated far more strongly than a typical MSP-WD binary, and nearly as strongly as the RBs. However, because WD binaries are nine times more numerous than RBs, the weighted likelihoods of the two classes are comparable, and hence not definitive on their own to distinguish between the two classes. Hence, the classification of the system, as discussed in Section~\ref{sec:discussion}, is supported and relies on other constraints such as the companion mass, Roche-lobe analysis and eclipse properties.

\section{Frequency evolution of the profile components}\label{components}

\begin{figure*}
    \centering
    \includegraphics[width=0.4\textwidth]{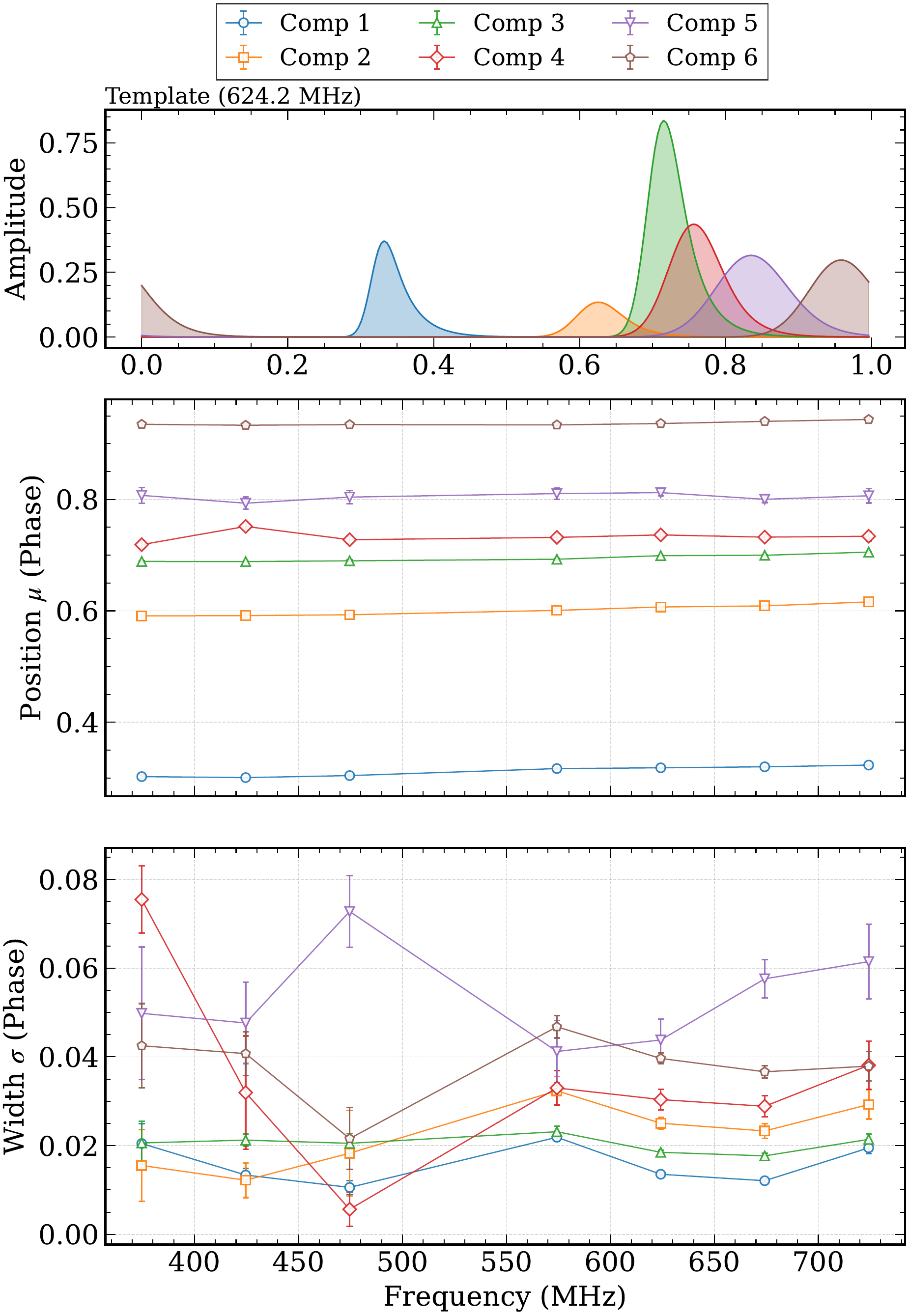}
    \caption{Frequency evolution of the intrinsic pulse profile components for PSR J1840+1102. \textbf{Top:} The reference exponential-convolved Gaussian component template derived from the high-SNR sub-band at 624.2\,MHz. \textbf{Middle and Bottom:} The fitted centroid phase positions ($\mu$) and a representative of characteristic widths ($\sigma$) of each component tracked across all valid sub-bands. The data points are color-coded to match the shaded components in the top panel.}
    \label{fig:pos_wid_evo}
\end{figure*}

Figure~\ref{fig:pos_wid_evo} shows the six Gaussian components used to model the intrinsic pulse profile in Section~\ref{scattering}, together with their fitted positions ($\mu$) and representative widths ($\sigma$) in the seven sub-bands between 374 and 724 MHz. The values are the medians of the MCMC posterior distributions, with error bars marking the 16th and 84th percentiles. The lowest-frequency sub-band of band-3 is not included because of its low SNR. 

\pagebreak

\bibliography{Timing-2,manual}{}
\bibliographystyle{aasjournalv7}


\end{document}